\documentclass{article}

\usepackage{arxiv}

\usepackage[utf8]{inputenc} 
\usepackage[T1]{fontenc}    
\usepackage[hidelinks]{hyperref}       
\usepackage{amsmath}
\usepackage{url}            
\usepackage{booktabs}       
\usepackage{amsfonts}       
\usepackage{nicefrac}       
\usepackage{microtype}      
\usepackage{lipsum}
\usepackage{graphicx}
\usepackage[backend=biber, style = authoryear]{biblatex}
\usepackage{placeins}
\usepackage{xcolor}
\usepackage{booktabs}
\usepackage{makecell}
\usepackage{tcolorbox}
\usepackage{caption}


\usepackage{bm}

\title{Leveraging Large Language Models to Improve
Precision in Randomized Controlled Trials}

\author{
 Jaylin Lowe \\
  University of Michigan\\
  Ann Arbor, MI, 48104 \\
  \texttt{jaylincl@umich.edu} \\
   \And
Adam Sales \\
  Worcester Polytechnic Institute\\
  Worcester, MA 01609 \\
  \texttt{asales@wpi.edu} \\
  \And
Johann A. Gagnon-Bartsch \\
  University of Michigan\\
  Ann Arbor, MI, 48104 \\
  \texttt{johanngb@umich.edu} \\
}

\begin{document}
\maketitle
\begin{abstract}
Large language models (LLMs) are increasingly used in statistical research and applications. However,
they are also notorious for unreliable or biased information. Here, we explore whether LLMs can be
used to improve the precision of randomized controlled trials (RCTs) in a safe and rigorous way. We
consider two pathways for incorporating LLM assistance: (1) using LLMs to transform non-standard
covariates, such as text, into numerical representations, and (2) leveraging the general knowledge
embedded in LLMs to generate predicted outcomes. Although external data sources have been
leveraged to improve precision before, the value added by LLM-generated predictions through these
pathways remains an open question. We develop a pipeline for generally leveraging LLMs in this context and apply it to three different case studies. We find that we are able to safely improve precision, primarily through transforming non-standard covariates, but that LLMs also possess general
knowledge that is useful in some contexts.

\end{abstract}

\section*{Introduction}

Large language models (LLMs) are now widely used for a variety of tasks. However, many researchers have expressed concerns that LLMs hallucinate, respond inconsistently to prompts, and amplify training data biases \parencite{majeed_reliability_2024, ye_assessing_2023, mahaut_factual_2024, zheng_how_2024}. In this paper, we investigate whether LLMs can be used to improve precision in randomized controlled trials (RCTs) in a safe and rigorous way. We focus on how to do this in practice.

RCTs have long been considered the ``gold standard'' in causal inference. However, RCTs are often small, especially in the social sciences, leading to estimates with high variance. This is often addressed via covariate adjustment, discussed more in the next section. Covariates are measured characteristics recorded for each observation that may be associated with the outcome; covariate adjustment models small chance imbalances in these characteristics across groups \parencite{fisher_statistical_1925}. Traditionally, covariate adjustment models are fit on just the RCT data. One problem with this approach is that the covariates and outcomes may be hard to model. For instance, the covariates could be high-dimensional, have a complex structure, or be in a form, like text, that is not compatible with standard statistical models. In such cases, the RCT sample size often is not large enough on its own to estimate a predictive model.

LLMs may help address these issues in two main ways. First, LLMs can be used to handle non-standard covariates, such as text. LLMs can process these covariates in their original forms and transform them into numerical representations. Second, LLMs can leverage their “general knowledge”—information obtained via their extremely large training datasets—to make predictions of the outcome. Both approaches produce numerical output—representations of non-standard covariates in one case and predicted outcomes in the other—that can be treated as conventional covariates suitable for effect estimators. We incorporate covariates into treatment effect estimation using an approach that guarantees unbiasedness without compromising precision in moderate to large samples. Thus, we need not rely on LLM accuracy to ensure unbiasedness, nor does LLM inaccuracy compromise precision.
 
In principle, using an LLM generated covariate in a RCT analysis is fairly straightforward. However, determining the best methods to construct these covariates is an open question. In this paper, we present a pipeline for constructing LLM-generated covariates and apply it to three different case studies. For the first—an RCT estimating the effect of open access on citation counts for scientific papers—we obtained the papers’ abstracts and incorporated them as covariates. This case study allows us to examine how LLMs can both process textual covariates and draw on their general knowledge to predict which papers will receive more citations. In this case study, our method substantially improves precision, with one setting achieving gains comparable to increasing the sample size by nearly 50\%. To explore the benefit of the LLM’s general knowledge apart from its ability to transform nonstandard covariates, we also examine two additional case studies. These case studies were specifically chosen to feature only rectangular, quantitative covariates and outcomes for which a LLM may reasonably possess relevant general knowledge. Our results from these case studies indicate that the LLM can leverage their training data to produce informative predictions of outcomes.

This paper is organized as follows. First, we introduce the relevant causal inference notation and provide background on related work for improving precision in RCTs. Next, we introduce the three case studies. Our Methods section gives our approach for obtaining LLM predictions and evaluating their utility. Then, we apply the method to each of the studies, report the results, and discuss our overall findings. In our Limitations section, we discuss a potential threat to the method’s validity and provide evidence that it does not appear to be a concern in our specific case studies. The final section, Discussion, concludes.

\section*{Background}

\subsection*{Randomized Controlled Trials}

We will use the potential outcomes framework developed by Neyman and Rubin \parencite{rubin_causal_2005, splawa-neyman_application_1990}. Under this framework, we have $N$ observations, indexed by $i = 1, .... N$. Each observation is randomly assigned to treatment ($Z_i = 1$) or control ($Z_i = 0$), where $Z$ is the vector of treatment assignments. Consider a Bernoulli experiment, where all units are independently randomized with $P(Z_i = 1) = p$. We have a vector of RCT covariates $\mathbf{x}_i$ for each observation. Let $n_t$ and $n_c$ represent the number of observations assigned to treatment and control, respectively. Each observation has a potential outcome under treatment, $y_i^t$, and a potential outcome under control, $y_i^c$. The potential outcomes $y_i^t$ and $y_i^c$ represent the outcome value we would observe if observation $i$ were assigned to treatment or control, respectively. The individual treatment effect $\tau_i$ is defined as $y_i^t - y_i^c$. We want to estimate the average treatment effect:

\begin{align} \label{tau_bar}
    \bar{\tau} = \frac{1}{N} \sum_{i=1}^N \tau_i = \frac{1}{N} \sum_{i=1}^N (y_i^t - y_i^c)
\end{align}

We only observe $y_i^t$ or $y_i^c$ for each observation, but never both. Thus, $\bar{\tau}$ is not directly calculable, but may be estimated. Consider the Horvitz-Thompson estimator \parencite{horvitz_generalization_1952, aronow_class_2013}, which estimates the average treatment effect as:  

\begin{align} \label{ipw}
    \hat{\tau} = \frac{1}{N} \sum_{i=1}^N Z_i \cdot \frac{Y_i}{p} - \frac{1}{N} \sum_{i=1}^N (1 - Z_i) \cdot \frac{Y_i}{1-p}
\end{align}

That is, we estimate the average treatment effect as the difference between the average of the outcome in the treatment group and the average of the outcome in the control group, weighted by the assignment probability. In the next section, we discuss alternatives to this estimator that help reduce the variance of our estimate. 

\subsection*{Improving Precision in RCTs}

\noindent Random assignment to the treatment and control groups ensures that all covariates will be balanced between the groups in expectation. However, covariates can still be imbalanced between the groups by chance, especially in smaller samples, leading to high variance. Covariate adjustment addresses this imbalance by leveraging the RCT covariates to account for chance imbalances, leading to a more precise estimator.

One simple option for covariate adjustment is to run a linear model regressing the outcome on the treatment indicator and the covariates \parencite{fisher_statistical_1925, freedman_regression_2008}. Others have proposed improvements to this general framework \parencite{lin_agnostic_2013}. An alternative approach is model imputation, in which observed outcomes and covariates are used to impute each observation’s missing potential outcome—$y_{i}^t$ or $y_{i}^c$. Traditionally, these approaches train models to predict treatment and control potential outcomes separately \parencite{oaxaca_male-female_1973, blinder_wage_1973}. In the simplest versions of this class of estimators, the imputations are based on models trained on the RCT covariates. More recent approaches extend this framework by integrating RCT data with larger external datasets to improve precision \parencite{deng_improving_2013, gagnon-bartsch_precise_2023-1, gui_combining_2024, kallus_removing_2018, rosenman_combining_2023, rosenman_propensity_2022}. Large external datasets can be used to produce predicted outcomes. These predictions can then be included in the model as an additional univariate covariate. Conceptually, our approach builds upon this strategy by using LLMs, rather than models trained on large external datasets, to generate the predicted outcomes. LLMs are trained on extremely large training sets, which may even include relevant external data sources. Thus, using an LLM to produce predicted outcomes can be thought of an alternative, indirect way of using external datasets to generate predicted outcomes, except we do not need to be able to access the external datasets ourselves.

These LLM-generated numerical representations or predicted outcomes can be treated as auxiliary covariates in a broad class of effect estimators. We chose to use the approach outlined in \cite{gagnon-bartsch_precise_2023-1}, as our estimator, as it is flexible and has good estimates of the standard error. This method is design-based and makes no assumptions about the quality of the predictions. The method, adapted to use LLM predictions, is as follows:
\begin{enumerate}
    \item Obtain a vector of predictions $\hat{y}_i^{LLM}$ from an LLM.  These predictions will be used as a covariate in the future, so let $x_i^{LLM} \equiv \hat{y}_i^{LLM}$. 
    \item Augment the vector of RCT covariates, $\mathbf{x}_i$, with this additional covariate. We call the resulting vector $\mathbf{\tilde{x}}_i$ where $\mathbf{\tilde{x}}_i = \begin{bmatrix} \mathbf{x}_i, x_i^{LLM} \end{bmatrix}$
    \item Obtain imputations $\hat{y}^t (\mathbf{\tilde{x}}_i)$ and $\hat{y}^c (\mathbf{\tilde{x}}_i)$ for each observation $i$ in the RCT. Crucially, the imputations for observation $i$ must be independent of observation $i$'s treatment assignment. These could be leave-one-out predictions from a linear model, out-of-bag predictions from a random forest, or something similar. This step allows for flexible use of $x_i^{LLM}$. If it is useful, then each $\hat{y}(\mathbf{\tilde{x}}_i)$ will take advantage of it. If not, each $\hat{y}(\mathbf{\tilde{x}}_i)$ will be based on the RCT covariates. 
    \item Calculate \begin{align}\hat{\tau} = \frac{1}{N} \sum_{i = 1}^n Z_i * \frac{Y_i - \hat{m}_i}{p} - \frac{1}{N} \sum_{i = 1}^n (1 - Z_i) * \frac{Y_i - \hat{m}_i}{1 - p} \end{align} where 
    $\hat{m}_i = p \hat{y}^c (\mathbf{\tilde{x}}_i) + (1 - p) \hat{y}^t (\mathbf{\tilde{x}}_i)$. This is our estimated average treatment effect. Note that this has the same form as the AIPW estimator \parencite{robins_aipw} and other related methods \parencite{aronow_class_2013,tsiatis_covariate_2008,wager_high-dimensional_2016}. 
    \item Calculate the mean squared prediction error in the control group ($\hat{E}_c^2$) and the treatment group ($\hat{E}_t^2$):
    \begin{align}
        \hat{E}_c^2 = \frac{1}{n_c} \sum_{i = 1}^N (1-Z_i) * [y_i^c-\hat{y}^c_{-i}(\mathbf{\tilde{x}}_i)]^2 \text{ and } \hat{E}_t^2 = \frac{1}{n_t} \sum_{i = 1}^N Z_i * [y_i^t-\hat{y}^t_{-i}(\mathbf{\tilde{x}}_i)]^2
    \end{align}
    Then, the estimated variance of $\hat{\tau}$ is:

\begin{align} \label{var_eq} \widehat{\mathrm{Var}}(\hat{\tau}) = \frac{1}{N} \left[ \frac{p}{1-p} \hat{E}_c^2 + \frac{1-p}{p} \hat{E}^2_t + 2 \sqrt{ \hat{E}^2_c \hat{E}^2_t} \right] \end{align}
    
\end{enumerate}
Since the LLM prediction augments the data already collected in the RCT, a $\hat{y}(\mathbf{\tilde{x}}_i)$ obtained using $\mathbf{\tilde{x}}_i$ may contain more information than a version based on $\mathbf{x}_i$ alone. If $\hat{y}(\mathbf{\tilde{x}}_i)$ is highly correlated with the RCT outcome, then significant gains in precision can be made.

\section*{Case Studies} 

\subsection*{Case Study 1: Open Access Paper Citations}

Our first case study is an RCT aimed at investigating the impact of open access policies on paper citations \parencite{davis_open_2011}. A total of 3,245 papers published in 2007 across 36 different journals were randomly assigned to either the treatment or control group. Papers randomized to treatment were made open access upon publication. Papers randomized to control were only available to subscribers, subject to the journal’s typical policy. The outcome of interest is the number of citations each paper receives 3 years after publication, log transformed. Only four RCT covariates were originally available.

While the original RCT contained papers from 36 journals, we restrict to journals with at least 150 papers in the RCT where we could obtain abstracts from the PubMed API. This approach ensured there were enough papers from each journal to conduct separate analyses. We use 1,248 papers across five journals. Table \ref{tab:paper_counts} displays the journals, sample sizes, proportion in treatment and control, and the length of delay until all papers in that journal are open access. The proportion of journals assigned to treatment varies by journal; this was determined in collaboration with the publishers of the journals and depended on size, frequency, and a desire to avoid significant changes in appearance or performance of the journal.

\begin{table}[]
\centering
\small
\caption{Sample Sizes and Delayed Access Policy by Journal}
    \label{tab:paper_counts}
 \begin{tabular}{ p{4.5cm}  p{2 cm} p{2 cm}   p{2 cm}  p{2.5cm}} 
 \hline
 Journal   & N & Treatment &  Control & Delayed Access \\ [0.5ex] 
\hline
 Science & 393 & 0.122 & 0.878  & 12 mo \\ 
 Journal of Neurophysiology & 278  & 0.140 & 0.860   & 12 mo \\
 Genetics & $211^{*}$ & 0.488 & 0.512  & 6 mo\\
 Journal of Applied Physiology & 201 & 0.134 & 0.866  &  12 mo\\
 FASEB & $165^{*}$ & 0.491 & 0.510  & 12 mo\\ [1ex] 
 \hline
 \end{tabular}
 
 \vspace{0.5em}
\footnotesize{
Sample sizes are reported as proportions in treatment and control for each journal. Journals with an asterisk (*) had exactly one paper published without an abstract. These papers were removed. }
\label{oa_cov}
\end{table} 

\subsection*{Case Study 2: Sentencing of Defendants and Recidivism}
Our second case study is a natural experiment investigating the effects of incarceration and probation on recidivism \parencite{green_winik_2010}. This study takes advantage of the random assignment of judges to defendants. Since different judges naturally have varying predispositions toward shorter or longer sentence lengths, random assignment can be leveraged to assess the impact of sentences via an instrumental variable analysis. We instead focus on the direct effect of judge assignment on case outcomes, considering judge assignment as the treatment of interest. The case study contains demographic information along with data on past arrests and convictions for 1,003 defendants arrested on drug-related charges in the District of Columbia in 2002 and 2003. The outcome of interest, recidivism, is a binary variable capturing whether the defendant was arrested again within four years. A table giving the summary statistics of the covariates can be found in Appendix \ref{sec:gw_appendix}. 

\subsection*{Case Study 3: Cognitive Tutor Algebra (CTA)}

Our third case study is an efficacy trial for a new proposed algebra curriculum called the Cognitive Tutor Algebra I curriculum \parencite{pane_effectiveness_2014}. This RCT was cluster pair randomized; schools were paired and randomized to either continue with their original algebra curriculum (control) or switch to the CTA curriculum (treatment). Our data is at
the student level and includes demographic covariates for 19,053 students in addition to each student’s standardized score on an algebra readiness exam administered before the experiment. A table of summary statistics can be found in Appendix \ref{sec:cta_appendix}. 

\section*{Methods}

\subsection*{Overview}

Our goal is to obtain an additional covariate from the LLM that is predictive of the RCT outcome. The most straightforward approach is to ask the LLM to predict each observation’s outcome. However, this method tends to predict overly similar results for each observation. For example, for the recidivism case study, we initially asked the LLM to predict whether each defendant will re-offend. It predicted that none of the defendants in the case study would re-offend.

Accordingly, rather than generating a separate prediction for each observation, we pair observations and obtain pairwise comparisons from the LLM. For example, instead of asking the LLM to predict whether each defendant will re-offend, we give the LLM pairs of defendants and ask it to predict which defendant is more likely to re-offend. We can convert these pairwise comparisons into a count for each observation of the number of times it was chosen over its partner. For the recidivism case study, this count can intuitively be thought of as a “risk score” of how risky a particular defendant is considered compared to the other defendants in the case study. In the other case studies, this “score” is not as interpretable. Regardless, these “scores” can be used as an additional covariate that may be predictive of the RCT outcome. Our method consists of three stages: question formulation; covariate construction and stratification; and modeling and evaluation. We discuss each step in detail

\subsection*{Question Formulation}
All our questions follow a similar format. First, we state that we have information about two observations, which can be papers, defendants, or students, depending on the case study. Then, we give all the covariates for the first observation, followed by the same for the second observation, written out in sentence form. If there are observations with missing values, we simply tell the LLM that the covariate is unknown rather than omitting it.

The LLM is asked to predict which observation is more likely to exhibit a specific quality. Usually, we instruct it to respond with either “observation 1” or “observation 2”, comparing the two on the outcome variable. For the first case study, we also asked the LLM to compare observations according to 10 additional qualities, such as which paper is more creative, has a more popular topic, a more meaningful contribution or similar. The prompts given to the LLM for all three case studies can be found in Appendix \ref{sec:prompt_appendix}.

Occasionally, the LLM may refuse to answer. Sometimes this happens when both observations in a pair have extremely similar values. In general, refusal to answer was not a common issue. We discovered that if the LLM does refuse to answer, often it will answer the exact same question repeated a second time. We did not observe cases where the LLM refused to answer a large proportion or all the questions.

\subsection*{Covariate Construction and Stratification}

In this step, we use the question or questions chosen in the previous step to construct one or more LLM-generated covariates. Each question asks the LLM to make a comparison between a pair of observations. For each pair of observations $(i,j)$, we give the LLM two separate queries: one where observation $i$ is presented first and one where observation $j$ is presented first. This is necessary to mitigate order effects, as we initially observed that the LLM exhibits a slight preference for the observation presented first. We will have two predictions for each unique pair of observations. For each observation, we count the number of times it was chosen over its partner. If the LLM gave any response other than “Observation 1” or “Observation 2”, we dropped that pair. If the number of comparisons performed was not equal for all observations—such as if comparisons were dropped—we divide by the number of comparisons performed for that group to maintain consistency. We call the resulting value the adjusted pair score. For the first case study, there are 11 adjusted pair scores: 1 for the comparison about the outcome variable (citation count) and 10 additional ones for each additional quality we asked about (creativity, etc.). For the other two case studies, there is only one adjusted pair score. These adjusted pair scores are now additional covariates that can be used to improve precision.

This step also requires deciding whether to compare every possible pair of observations or only a selected subset. In some situations, using a subset may be more appropriate. For example, in our first case study, we stratify by journal and look only at pairs of papers where both papers were published in the same journal. Different journals tend to have very different citation counts, so papers from different journals may not be directly comparable. In our second case study, an initial random forest suggested that age was the most important covariate. Since the goal is for the LLM to provide information beyond what traditional methods already capture, comparing defendants who are similar in age would have been one reasonable choice.

Instead of choosing a specific covariate, a more general option is to fit an initial model on the RCT covariates and stratify by its predictions. Since our goal is to obtain information not captured by models fit to the RCT data, stratifying on predictions from an initial model requires the LLM to distinguish observations beyond what a model fit to the RCT already can. For our second and third case studies, we stratify on out-of-bag predictions from a random forest. These random forests are fit on all observations and with the default parameters, as default parameters are known to perform well for random forests \parencite{breiman_rf_2001, fernandez_classifiers_2014}.

Beyond the gains in precision, stratification can also substantially reduce computational time. Without stratification, our second and third case studies would require over a million ordered pairs and over 360 million ordered pairs, respectively. We stratify the second case study into groups of size 100, reducing the number of pairs to 99,606. For the third case study, we stratify into groups of size 10, resulting in 171,456 pairs for analysis.

\subsection*{Modeling and Evaluation}

To see if the LLM predictions were useful, we check the significance of the LLM-generated covariate in a regression model fit to the original RCT covariates plus the new LLM-generated covariate. Additionally, we compare the variance of the estimator with the LLM-generated covariate to that without. We compare the ratio of these variances to obtain and estimate of the effective sample size increase. The variance of the estimator is calculated using Equation \ref{var_eq}\footnote{For the recidivism case study, there are nine treatment categories, so Equation 5 cannot be applied directly. Instead, we calculate pairwise standard errors for all possible pairs of judges. For example, we calculate the standard error of an estimator comparing defendants assigned Judge 1 to defendants assigned Judge 2.}.

\subsection*{Implementation Details}
 We use OpenAI’s “gpt-4o-mini” model as our LLM, because OpenAI models are among the most widely used in practice. “Gpt-4o-mini” is one of OpenAI’s smaller and older models, so it is cheaper and easier to run many queries.  All code, AI prompts, and public datasets are available at \href{https://github.com/jaylinlowe/llm-rct}{github.com/jaylinlowe/llm-rct}.

Running our analysis for case study 1 takes approximately 150 hours and costs \$70.00. This cost varies substantially by journal: \$22.90 for Science, \$18.48 for the Journal of Neurophysiology, \$10.88 for the Journal of Applied Physiology, \$9.08 for Genetics, and \$6.44 for FASEB. This encompasses running two queries for each ordered pair: one that asks the LLM to predict the paper with more citations, and one that asks the LLM to predict which paper best exhibits each one of 10 different qualities. Journals with more papers in the RCT are substantially more expensive. For case study 2, it takes approximately 26 hours and costs \$6.00. For case study 3, it takes approximately 20 hours and costs \$7.00. In general, the first case study is much more expensive per query because each query includes two abstracts, versus simpler covariates in the other RCTs.

\section*{Results}

\subsection*{Open Access Paper Study}

\cite{davis_open_2011} employed four covariates: number of authors, number of papers, whether the paper was self-archived or not, and whether the paper was a review paper. We use the titles and abstracts. Since citations can vary widely by journal, we stratify by journal. Results are presented for each journal separately. For each pair of articles, we give the LLM the titles and abstracts of both papers. Each ordered pair is given to the LLM twice. The first time, the LLM is asked to predict which paper in the pair will have the higher citation count. The second time, the LLM is given a list of 10 qualities and asked to list which paper exhibits each quality best. The 10 qualities are: topic novelty, topic popularity, title catchiness, generalizability, writing quality, impact of results, subfield popularity, technicality, meaningful contributions, and applicability. 

After querying the LLM, we have 11 pair scores: one for the comparison about the number of citations and 10 additional ones for each additional quality we asked about. An analysis of these 11 scores, including their means, standard deviations, correlations with the outcome, and histograms can be found in Appendix \ref{sec:oa_appendix}. We fit four random forest models with different covariates to predict log citation counts:

\begin{enumerate}
    \item Model 1: includes only the four base covariates. 
    \item Model 2: includes all the base covariates plus our citation score.
    \item Model 3: includes all the base covariates plus our 10 quality scores.
    \item Model 4: Includes the base covariates, citation score, and 10 quality scores.
\end{enumerate}

For each observation, a version of each of these four models is fit with that observation dropped from the analysis. Each model also includes the treatment variable. The resulting models are used to make predictions $\hat{y}^c(\mathbf{\tilde{x}_i})$ and $\hat{y}^t(\mathbf{\tilde{x}_i})$ for each observation. These imputations are used to calculate an average treatment effect (Equation \ref{ipw}) and its estimated variance (Equation \ref{var_eq}). Thus, we have four estimators for each journal, one for each model type. Additional information about these models and the estimator standard errors can be found in Appendix \ref{sec:oa_appendix}. Figure \ref{fig:oa_results} gives the effective sample size percentage increase for each journal and model type, relative to the base model (model 1). This is calculated as the ratio of the variance of the estimator with the base covariates compared to the variance of the estimator with some combination of LLM-generated covariates, converted to a percentage. Values above 0 indicate a more precise estimator. For example, if we used all 11 LLM-generated covariates for Science, this would be comparable to increasing the sample size by nearly 50 \% (upper leftmost green square).

\begin{figure}[ht!]
    \centering
    \includegraphics[width=0.7\textwidth]{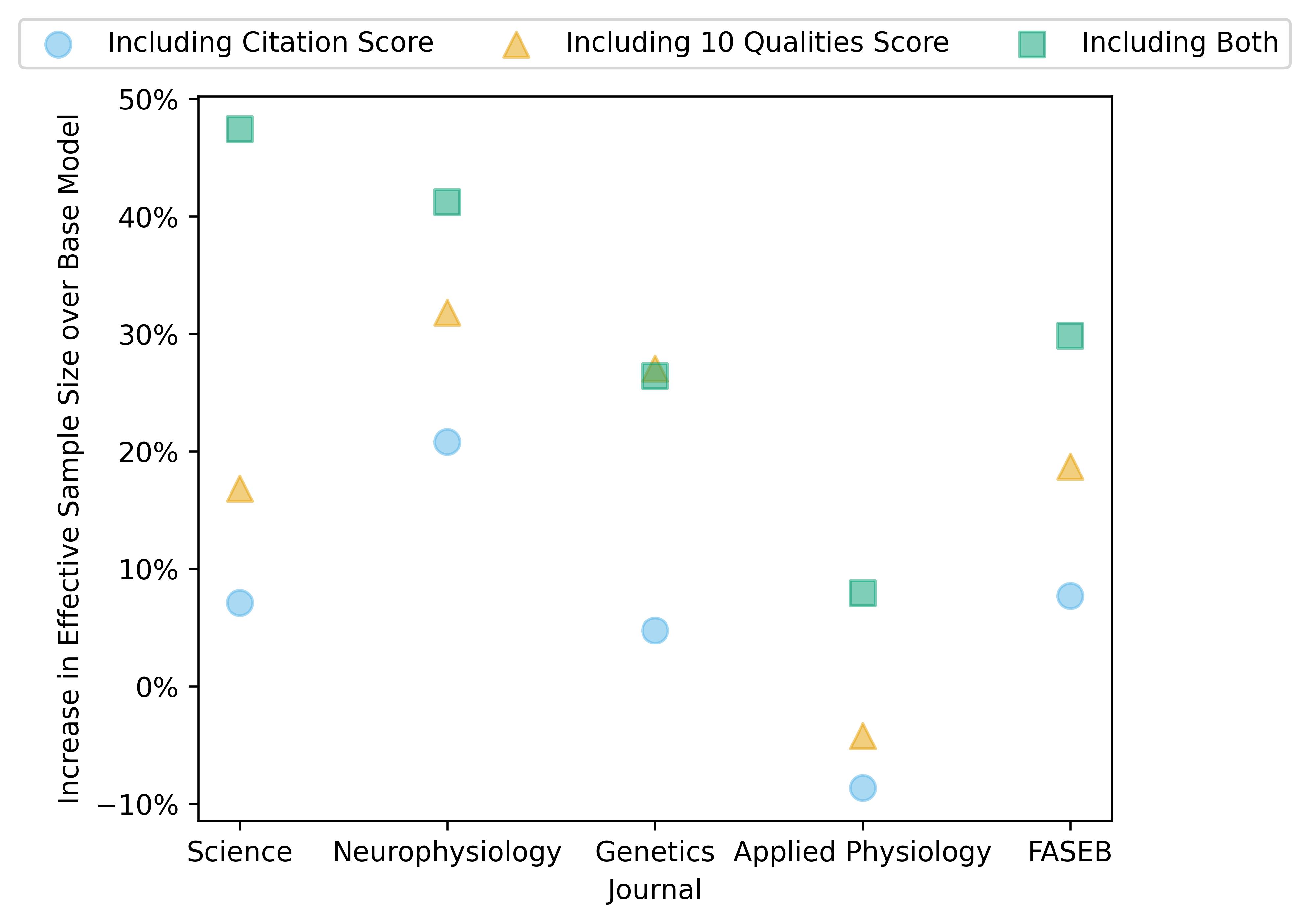}
    \caption{Equivalent Sample Size Improvement, by Model Type and Journal}
    \label{fig:oa_results}
    \vspace{0.5em}
    
\end{figure}

As desired, we see that precision improves when we add in the LLM-derived covariates. The degree of improvement and comparisons between the citation score alone and the quality scores vary by journal, but all journals see an improvement in the standard error of the estimator when LLM predictions are included. The gains in precision are largest when the citation score and 10 quality scores are used together (green squares). This suggests that the LLM adds value in both ways: (1) by transforming nonstandard covariates and (2) by incorporating general knowledge. The value of the quality scores—numerical representations of the abstracts—supports the first mechanism. The value of the citation scores—which reflect the LLM’s general knowledge of highly cited papers—supports the second.

\subsection*{Sentencing of Defendants and Recidivism}

Our second case study examines whether the LLM can add value only through leveraging its general knowledge. Because the original RCT covariates are standard rectangular covariates, there is little opportunity for the LLM to add value by transforming them. However, recidivism is a widely studied outcome that the LLM likely encountered in its training data, making its general knowledge potentially valuable.

We first stratify defendants into 10 groups of approximately equal size, based on out of bag predictions from a random forest trained on the original covariates. The adjusted pair score is not statistically significant in a logistic regression model (Table \ref{tab:gw_regress_results} in Appendix \ref{sec:gw_appendix}). This model includes all the RCT covariates, the adjusted pair score, and the out of bag predictions used to stratify the observations. Thus, the LLM predictions are not useful in explaining any of the variance of the outcome not already explained by the other RCT covariates. Thus, we do not expect to see an improvement in the variance of the treatment effect estimate, but we still calculate the variance of the estimator to demonstrate that our approach does not hurt precision. Table \ref{tab:gw_var_results} in Appendix \ref{sec:gw_appendix} gives the estimator standard errors for pairwise comparisons of defendants assigned to different judges. The standard errors with and without the adjusted pair score are very similar (see Figure \ref{fig:gw_scatter} in Appendix \ref{sec:gw_appendix}).

Here, we also experimented with requesting a short one-sentence explanation from the LLM. These explanations reveal that the gpt-4o-mini tends to prioritize a defendant’s history—past arrests or convictions—over other factors when determining who is most likely to re-offend. This pattern is further supported by our analysis of the adjusted pair score, which shows a stronger correlation with prior-history variables than with any other variables (see Table \ref{tab:gw_cov_corr} in Appendix \ref{sec:gw_appendix}). Indeed, there is evidence that prior history does predict recidivism, especially in the short periods following an offense (\parencite{kurlychek_scarlet_2006, gendreau_meta-analysis_1996}. However, in this case, prior history does not predict recidivism. In general, none of the original covariates are notably predictive. Age is the most predictive covariate, as assessed by variable importance, but is still a weak predictor overall (Table \ref{gw_varimp} in Appendix \ref{sec:gw_appendix}). Our adjusted pair score does not appear strongly affected by age—the correlation with age is small (0.277).

\subsection*{Cognitive Tutor Algebra}

Like our second case study, our third case study also has typical, rectangular covariates. This case study examines whether the LLM can add value through its general knowledge in an even more common setting: educational RCTs. Unlike our second case study, our third case study has many highly predictive covariates. It has a pretest covariate—scores on a similar exam before treatment—that is highly predictive. Due to the large sample size for this study, we stratify on the out of bag predictions from a random forest to form groups of size 10. The LLM is asked to predict the outcome variable: which student in each pair is more likely to score higher on an algebra proficiency exam. An analysis of the resulting adjusted pair score can be found in Appendix \ref{sec:cta_appendix} and includes summary statistics and the correlation of the adjusted pair score with all the covariates.

We start by comparing a regression model with the LLM predictions to a regression model without these predictions. Both models include all the RCT covariates and the out of bag predictions from the original random forest used to stratify the students. We see that the adjusted pair score is statistically significant (see TTable \ref{tab:cta_results} in Appendix \ref{sec:cta_appendix}). Thus, the LLM predictions are still able to add something to the model beyond what the RCT covariates can. In this case, this result is especially impressive because the other covariates are already highly predictive. In Appendix \ref{sec:cta_appendix}, we test whether the ranking process itself creates a useful covariate and find that it does not; the LLM is likely leveraging its general knowledge to produce useful predicted outcomes.

Although the adjusted pair score is significant in the linear model, this does not translate into a practical difference in the variance of the estimator. The calculated standard errors are 0.0733 and 0.0722 for the estimators without and with LLM predictions, respectively\footnote{As this RCT was pair cluster-randomized, we use the approach outlined in \cite{mann2026generalframeworkdesignbasedtreatment} for our variance calculations. The imputations are generated via random forest.}. This corresponds to an effective sample size increase of 3\%. Although the LLM did not significantly improve precision in this setting, the statistical significance of its predictions is promising for future work.

\section*{Limitations}

Our analysis has two key limitations. First, many of our decisions were made based on exploratory analyses without held-out training sets. In particular, case studies 2 and 3 were treated as test beds as we developed our approach. Although the ordering was unintentional, our first case study (open access) was analyzed last; as a result, our analysis was substantially informed by analytic decisions developed through our experience with case studies 2 and 3. For transparency, we include a discussion below of what decisions were based on exploratory analysis versus determined ahead of time.	

Second, given our use of an off-the-shelf LLM, there is a possibility that the LLM had prior exposure to the post-treatment outcomes. If so, the LLM could have “cheated” and based its predictions on the post-treatment outcomes. This is concerning because our estimator’s unbiasedness guarantee depends on the LLM predictions being independent of treatment assignment. We discuss below the plausibility of this issue and present evidence suggesting that the LLM is not leveraging post-treatment information.

\subsection*{Decisions Based on Exploratory Analysis}
Early exploratory analysis led to three key decisions: to use an OpenAI model, to pair observations, and to include both orderings of each pair, $(i,j)$ and $(j,i)$. Crucially, two of these decisions—to pair observations and to use an OpenAI model—were based solely on exploratory analyses on case studies 2 and 3, the less successful analyses. Our most successful example, the RCT on open access papers, was only ever performed with paired data and an OpenAI model.

Only one OpenAI model was ever considered. We initially selected a Qwen model from Hugging Face because it was open source but later switched to an OpenAI model because OpenAI models are more widely used. We chose our final OpenAI model based on cost and computational constraints, not via comparisons of model performance on our datasets. Our decision to include both orderings of each pair was made after we observed order effects for case study 1. This is computationally expensive; when it is infeasible, randomly ordering the observations within each pair is a reasonable alternative.

Many decisions were also pre-specified and made independently of the results. These included: the stratification choices, the 10 qualities queried for the first case study, and the number of journals included for the first case study. We initially chose to stratify observations due to computational constraints. Stratifying by journal was an obvious choice for case study 1. For case studies 2 and 3, we stratified on out-of-bag predictions from a random forest, using them to summarize all covariates rather than selecting a single covariate. The stratum sizes were chosen for computational reasons.

\subsection*{Exposure to Post-Treatment Outcomes}
Due to the nature of the data and our results, we do not believe gpt-4o-mini used post-treatment information to make its predictions for the second and third case studies. Our method did not result in gains in precision for the second case study; therefore, we believe it is unlikely RCT outcomes were used in the LLM predictions. The data for the third case study was private. However, our first case study may be a cause for concern. Our version of gpt-4o-mini cannot search the internet; thus, post-treatment information is only an issue if the LLM was exposed to the dataset during training and made use of this information during prediction. However, we do not know what is in OpenAI’s training data.

We conducted two small exploratory analyses to assess the possibility that the LLM was cheating by basing its predictions on post-treatment outcomes in our first case study. First, we look at the performance of the 11 different LLM-generated covariates—one citation-based score and 10 quality-based scores. If the LLM cheated, then some of these covariates may benefit from that cheating more so than others. For instance, queries about citations can obviously directly benefit from access to outcome-related information, while queries about more subjective qualities, such as writing quality, are less likely to benefit from such information. We compared the results with the single pair score—which asks about citations directly—to the results with the 10 quality scores. We see that the results from the 10 qualities outperform the citation-based pair score for all five journals. Since the citation-based pair score is the covariate most likely to benefit from cheating, the fact that the 10 quality-based pair scores outperform it offers encouraging evidence that the LLM did not rely on post-treatment information. We also look at which of the 10 qualities were the most useful, determined by variable importance in the random forest. The most useful covariates are writing quality, meaningful contributions, and impact of results. These qualities are more subjective and are not directly tied to popularity or novelty—the attributes that would be most likely to benefit from cheating.

Second, we ran a small experiment with the papers published in Science to explore if the LLM might have accessed the papers in its training data. Science is the largest journal, and the journal for which we saw the largest gains in precision. For each paper published in Science in the case study, we asked the LLM to suggest a journal where we should try to submit this paper. If the LLM were cheating, it might recommend submission to Science on the basis that it knows Science accepts similar papers, since Science accepted that exact paper. In an open-ended question where the LLM was just asked to suggest a journal, the LLM never suggested Science. When given an option between Science, Nature, PLOS One, and Proceedings of the National Academy of Sciences (PNAS), the LLM chose Science only 12 percent of the time.

Users of this method will need to evaluate the risk of post-treatment information on a case-by-case basis. Highly publicized datasets are of a particular concern. If the LLM predictions truly are a post-treatment covariate, then it may introduce some bias into the estimate.

\section*{Discussion}

In specific settings, our proposed pipeline is effective in reducing the standard error of the treatment effect estimate. However, its effectiveness can vary substantially depending on the context.

Our first case study, an RCT about the impact of paper accessibility on citations, provides the clearest evidence of our pipeline’s value. Here, the LLM predictions serve two purposes: (1) to transform abstracts into standardized covariates and (2) to incorporate the general knowledge LLMs possess about paper citations and qualities. For this case study, we can reduce variance significantly by including LLM-generated covariates. Even when LLM predictions of the primary outcome perform well, we can further improve precision by predicting additional qualities and combining these with the outcome predictions. Although the journals are similar, the size of the improvement differs across journals. This suggests our method’s performance can vary even in closely related contexts.

Our second and third case studies can only leverage LLMs in one way: by incorporating the LLM’s general knowledge. We intentionally chose case studies with outcome variables the LLM was likely to be familiar with—recidivism and test scores, with mixed results. Case study 2 (recidivism) is an example of when our pipeline is not effective. Although recidivism is a common outcome about which we expected the LLM to have general knowledge, it is notoriously difficult to predict (Dressel and Farid, 2018; Glazebrook, 2010; Jamil et al., 2023). The observed covariates are not very predictive, so the LLM lacks informative inputs to guide its predictions. In case study 3 (CTA), it is remarkable that the LLM predictions are still significant, even though this does not translate to a meaningful reduction in effective sample size. This case study had a pretest covariate which captured each student’s algebra ability prior to the curriculum in a format like the eventual outcome. Typically, utilizing pretest covariates in covariate adjustment methods can result in substantial gains. In general, the effectiveness of LLMs as a source of general knowledge for covariate adjustment may vary substantially across applications.

We believe our method will be most valuable for studies which are small, follow similar patterns as other studies in the same context, have few covariates, and include non-standard covariates such as texts or images. Small studies may benefit most from our method because traditional covariate adjustment methods rely on models trained on the RCT covariates. With a sufficiently large RCT dataset, a model trained directly on the observed covariates could recover patterns like those encoded in the LLM’s general knowledge, rendering the LLM predictions unnecessary. For example, the CTA study had a very large sample, and although the LLM-generated covariate was statistically significant, including it only led to a small reduction in variance.

This method may be more effective when applied to studies whose outcomes fall within the LLM’s general knowledge. We intentionally selected case studies 2 and 3 because they examined common outcomes—recidivism and test scores, respectively—that the LLM was likely to have encountered during training. However, a common outcome does not guarantee that this knowledge will be useful. In the recidivism case study, for example, the LLM relied on the general relationship between prior history and recidivism, which did not hold in our experimental sample. Thus, RCTs conducted in unusual settings or populations that depart from general patterns may be less likely to benefit from this method. Lastly, this method may be especially valuable for RCTs with text- or image-based covariates because the method can leverage these data types, whereas traditional covariate adjustment methods cannot. As the value of data grows and researchers gain greater capacity to collect diverse data types, RCT datasets containing text or images may become more common.

\newpage
\appendix{}

\section{Supplementary Information for Case Study 1}
\label{sec:oa_appendix}
\FloatBarrier

\FloatBarrier
\subsection{Analysis of Adjusted Pair Score for Case Study 1}
\FloatBarrier

\begin{figure}[htbp]
    \centering
    \includegraphics[width=0.8\textwidth]{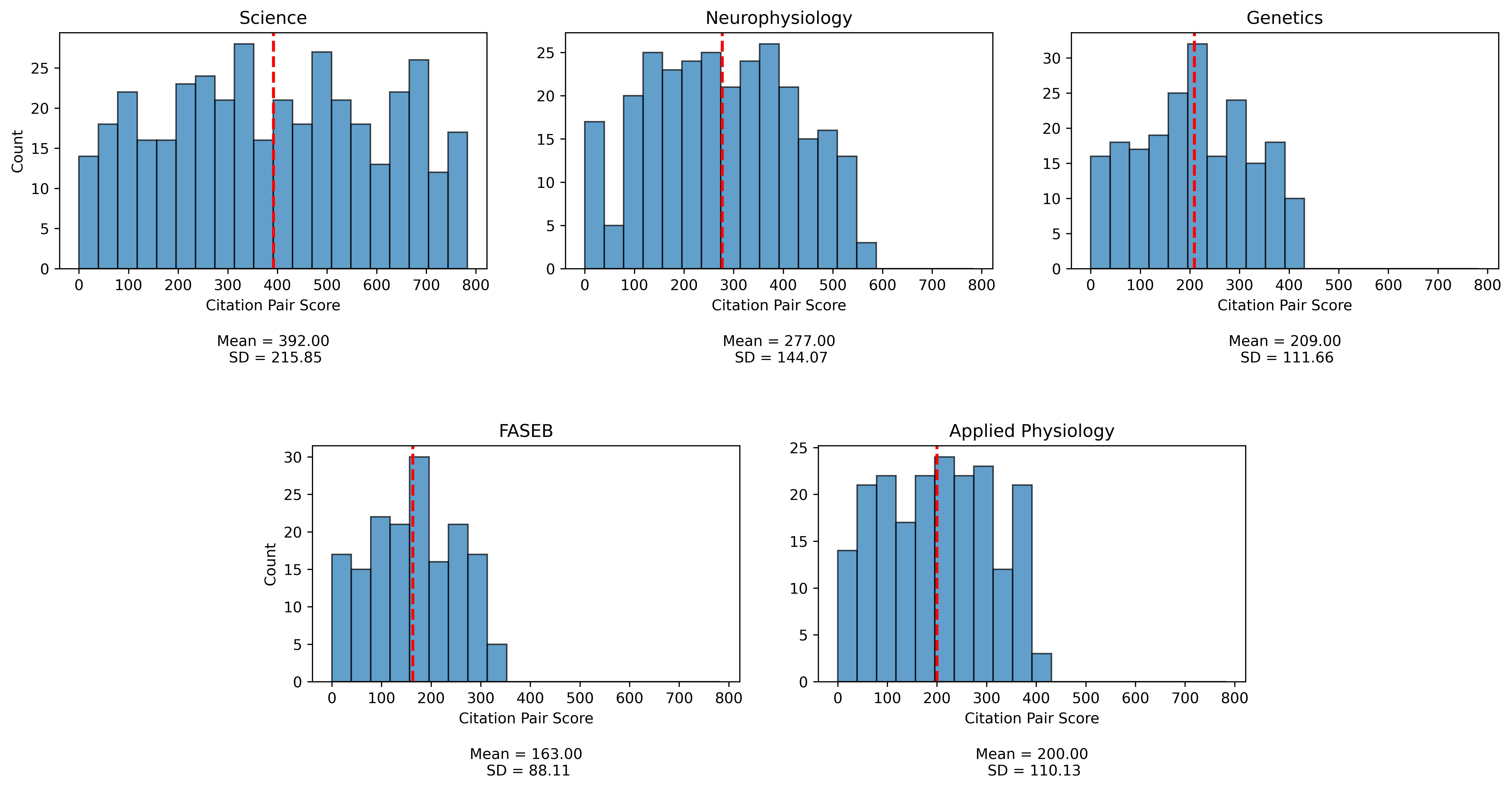}
    \captionsetup{justification=centering}
    \caption{Histogram of Adjusted Pair Score for Case Study 1}
    \label{fig:oa_hist}
\end{figure}

In this section, we provide an analysis of the adjusted pair score for Case Study 1. Figure \ref{fig:oa_hist} displays a histogram of the adjusted pair score by journal. The adjusted pair scores are not normalized by the number of comparisons—since no observations were dropped—so the magnitude of the pair scores differs slightly by journal. Table \ref{tab:correlations} displays the correlations of the adjusted pair scores with the outcome for each journal. We see that the basic citation-based pair score has fairly high correlations with the outcome, ranging from 0.341 to 0.512 across the journals. The different quality-based pair scores differ substantially in their correlation with the outcome. For example, impact of results and meaningful contributions have fairly high correlations, while subfield popularity and topic popularity have fairly low correlations.

\begin{table}[ht]
\caption{Correlation of Adjusted Scores with Outcome}
\label{tab:correlations}
\begin{tabular}{lrrrrr}
\toprule
 & Science & Neurophysiology & Genetics  & FASEB & Applied Physiology\\
Variable &  &  &  &  &  \\
\midrule
Basic Citations & 0.482 & 0.380 & 0.341 & 0.408 & 0.512 \\
Topic Popularity & 0.086 & 0.165 & 0.131 & -0.049 & 0.032 \\
Title Catchiness & 0.158 & 0.278 & 0.140 & -0.010 & 0.258 \\
Generalizability & 0.222 & 0.196 & 0.282 & 0.290 & 0.344 \\
Writing Quality & 0.389 & 0.337 & 0.034 & 0.117 & 0.326 \\
Impact of Results & 0.270 & 0.249 & 0.308 & 0.258 & 0.234 \\
Subfield Popularity & -0.140 & -0.055 & 0.059 & -0.148 & -0.004 \\
Technicality & -0.072 & -0.212 & -0.255 & -0.176 & -0.284 \\
Meaningful Contributions & 0.221 & 0.339 & 0.330 & 0.204 & 0.438 \\
Applicability & 0.248 & 0.115 & 0.287 & 0.175 & 0.132 \\
\bottomrule
\end{tabular}
\label{tab:oa_corrs}
\vspace{0.5em}

\footnotesize{The first row displays the correlation of the basic ``citation" pair score with the outcome for each of the five journals. The remaining rows show the correlation of each ``quality" pair score with the outcome.}
\end{table}

\FloatBarrier
\subsection{Additional Results for Case Study 1}
\FloatBarrier

In this section, we provide additional results for Case Study 1 that may be of interest to the reader. Although all our analyses for Case Study 1 are done at the journal level, we include a combined regression across all five journals for consistency with the other two case studies (Table \ref{tab:oa_combined_regression}). The pair scores are adjusted to account for the differing number of comparisons between each journal. We use separate random forests fit on each journal for our actual analysis (see Table \ref{tab: oa_rmse}) but include this table for consistency and to demonstrate that the basic citation score and some of the quality scores are significant. This regression suggests that all the LLM generated covariates are significant, except for topic popularity and impact of results. Table \ref{tab: oa_rmse} displays the RMSE for different random forests fit on all observations from a single journal. Our actual random forests are leave-one-out, but we include this table to give a benchmark of the performance of those random forests. Table \ref{tab:oa_ses} gives the estimator standard errors by journal. These values are used to calculate the ratios in Figure \ref{fig:oa_results} but we include the actual values here for reference. Table \ref{tab:oa_ps} reports the calculated p-values for each journal and model combinations. Only one is significant, which is to be expected for this number of p-values. 

\begin{table}[ht!]
\caption{Combined Regression Results for Case Study 1}
\centering
\begin{tabular}{lccc}
\toprule
 & Estimate & SE & p \\
\midrule
Intercept & 21.192 & 10.826 & 0.051 \\
Number of Authors (log) & 0.234 & 0.031 & 0.000 \\
Number of Pages (log) & 0.248 & 0.061 & 0.000 \\
Reveiw Article & 0.057 & 0.090 & 0.531 \\
Self-archived & 0.009 & 0.092 & 0.926 \\
Basic Citation Score & 1.052 & 0.115 & 0.000 \\
Topic Novelty Score & -45.050 & 22.889 & 0.049 \\
Topic Popularity Score & 0.930 & 3.229 & 0.773 \\
Title Catchiness Score  & -0.420 & 0.148 & 0.005 \\
Generalizability Score & 0.399 & 0.132 & 0.003 \\
Writing Quality Score & 0.668 & 0.190 & 0.000 \\
Impact of Results Score & -0.005 & 0.161 & 0.977 \\
Subfield Popularity Score & 1.195 & 0.455 & 0.009 \\
Technicality Score & 0.985 & 0.211 & 0.000 \\
Meaningful Contributions Score  & 0.649 & 0.269 & 0.016 \\
Applicability Score & 0.415 & 0.179 & 0.020 \\
Open Access & 0.048 & 0.046 & 0.299 \\
\textit{Journal} & \\
\quad{Genetics} & -0.437 & 0.069 & 0.000 \\
\quad{Applied Physiology} & -0.481 & 0.072 & 0.000 \\
\quad{Neurophysiology}. & -0.377 & 0.068 & 0.000 \\
\quad{Science} & 1.518 & 0.087 & 0.000 \\
\bottomrule
\end{tabular}
\label{tab:oa_combined_regression}
\end{table}

\begin{table}[ht!]
\caption{Random Forest RMSE by Journal and Model}
    \centering
  \begin{tabular}{lrrrr}
\\[-1.8ex]\hline
\hline \\[-1.8ex]
 Journal & Base Covariates & Base + Citation Score & Base + 10 Qualities & Base + Both \\
\midrule
Science & 0.8174 & 0.7624 & 0.7130 & 0.6863 \\
Neurophysiology & 0.7021 & 0.6287 & 0.6327 & 0.6108 \\
Genetics & 0.8081 & 0.7893 & 0.7278 & 0.7227 \\
Applied Physiology & 0.8293 & 0.7306 & 0.7226 & 0.6867 \\
FASEB & 0.6896 & 0.6597 & 0.6309 & 0.5972 \\
\\[-1.8ex]\hline
\hline \\[-1.8ex]
\end{tabular}
\label{tab: oa_rmse}
\vspace{0.5em}

\footnotesize{RMSEs are reported on the log scale. These are the RMSEs from random forests fit on all observations; in our analysis, we fit many leave-one-out random forests by iteratively dropping observations and fitting a random forest on the remaining data.
}
\end{table} 

\begin{table}[]
\caption{Estimator Standard Errors by Journal and Model}
    \centering
  \begin{tabular}{lrrrr}
\\[-1.8ex]\hline
\hline \\[-1.8ex]
 Journal & Base Covariates & Base + Citation Score & Base + 10 Qualities & Base + Both \\
\midrule
Science & 0.1227 & 0.1186 & 0.1136 & 0.1011 \\
Neurophysiology & 0.1300 & 0.1182 & 0.1132 & 0.1093 \\
Genetics & 0.1110 & 0.1085 & 0.0985 & 0.0988 \\
Applied Physiology & 0.1707 & 0.1786 & 0.1744 & 0.1643 \\
FASEB & 0.1066 & 0.1027 & 0.0979 & 0.0936 \\
\\[-1.8ex]\hline
\hline \\[-1.8ex]
\end{tabular}
\label{tab:oa_ses}
\end{table} 

\begin{table}[]
\caption{$p$ values by Journal and Model}
    \centering
  \begin{tabular}{lrrrr}
\\[-1.8ex]\hline
\hline \\[-1.8ex]
 Journal & Base Covariates & Base + Citation Score & Base + 10 Qualities & Base + Both \\
\midrule
Science & 0.152 & 0.008 & 0.286 & 0.111 \\
Neurophysiology & 0.423 & 0.409 & 0.642 & 0.523 \\
Genetics & 0.483 & 0.677 & 0.222 & 0.491 \\
Applied Physiology & 0.286 & 0.358 & 0.972 & 0.787 \\
FASEB & 0.288 & 0.099 & 0.137 & 0.145 \\
\\[-1.8ex]\hline
\hline \\[-1.8ex]
\end{tabular}
\label{tab:oa_ps}
\end{table}

\FloatBarrier
\clearpage
\section{Supplementary Information for Case Study 2}
\label{sec:gw_appendix}

\FloatBarrier
\subsection{Summary Statistics for Case Study 2}
\FloatBarrier

Here, we give basic summary statistics for Case Study 2. Table \ref{gw_cov1} displays the sample size and percentage of observations in each category for the categorical variables. Table \ref{gw_varimp} displays the variable importance for a random forest run on this dataset. As discussed in the paper, age is the most useful predictor. 

\begin{table}[ht]
\centering
\small
\caption{Summary statistics of covariates for Case Study 2}
\label{tab:covariates1}
\begin{tabular}{l c c}
\hline
Covariate & n & Percentage \\
\hline
\textit{Gender } &  \\
\quad{Female} & 115 & 11 \\
\quad{Male} & 888 & 89 \\
\textit{Race } &  \\
\quad{Black} & 25 & 2 \\
\quad{non-Black} & 978 & 98 \\
\textit{Prior History} & \\
\quad Prior Arrest &  856 & 85 \\
\quad Prior Drug Arrest &  724 & 72\\
\quad Prior Felony Arrest &  721 & 72\\
\quad Prior Drug Felony Arrest &  555 & 55\\
\quad Prior Conviction &  673 & 67\\
\quad Prior Drug Conviction &  551 & 55\\
\quad Prior Felony Conviction &  541 & 54\\
\quad Prior Drug Felony Conviction &  429 & 43\\
\textit{Charge} & \\
\quad Possession with intent to Distribute & 469 & 47 \\
\quad Distribution &  618 & 62\\
\quad{Additional Non Drug Charge} & 129 & 13 \\ 
\textit{Drugs Involved} & \\
\quad Cocaine &  401 & 40 \\
\quad Marijuana & 191 & 19 \\ 
\quad Crack & 197 & 20 \\
\quad Heroin & 264 & 26 \\
\quad PCP &  47 & 5 \\
\quad Other Drug &  33 & 3 \\[0.5em]
\hline
\end{tabular}

\vspace{0.5em}
\footnotesize{
N = 1,003. Defendants were on average 33.09 years old (SD: 11.25).
}
\label{gw_cov1}
\end{table}

\begin{table}[ht]
\centering
\small
\caption{Variable Importance of Covariates for Case Study 2}
\begin{tabular}{l c }
\hline
Covariate & Importance \\
\hline
Age & 0.504 \\
Additional Non Drug Charge & 0.039 \\
Drug: Cocaine & 0.038 \\
Charge: Possession with intent to Distribute & 0.038 \\
Gender & 0.035 \\
Charge: Distribution & 0.033 \\
Drug: Crack & 0.031 \\
Drug: Marijuana & 0.030 \\
Drug: Heroin & 0.030 \\
Prior History: Prior Felony Arrest & 0.027 \\
Pior History: Prior Felony Conviction & 0.025 \\
Prior History: Prior Drug Felony Arrest & 0.025 \\
Prior History: Prior Drug Arrest & 0.023 \\
Prior History: Prior Conviction & 0.023 \\
Prior History: Prior Drug Conviction & 0.022 \\
Prior History: Prior Arrest & 0.018 \\
Prior History: Prior Felony Drug Conviction & 0.017 \\
Drug: PCP & 0.016 \\
Race & 0.014 \\
Drug: Other Drug & 0.013 
 \\[0.5em]
\hline
\end{tabular}
\vspace{0.5em}
\label{gw_varimp}
\end{table}

\FloatBarrier
\subsection{Analysis of Adjusted Pair Score for Case Study 2}
\FloatBarrier

\begin{figure}[htbp]
    \centering
    \includegraphics[width=0.8\textwidth]{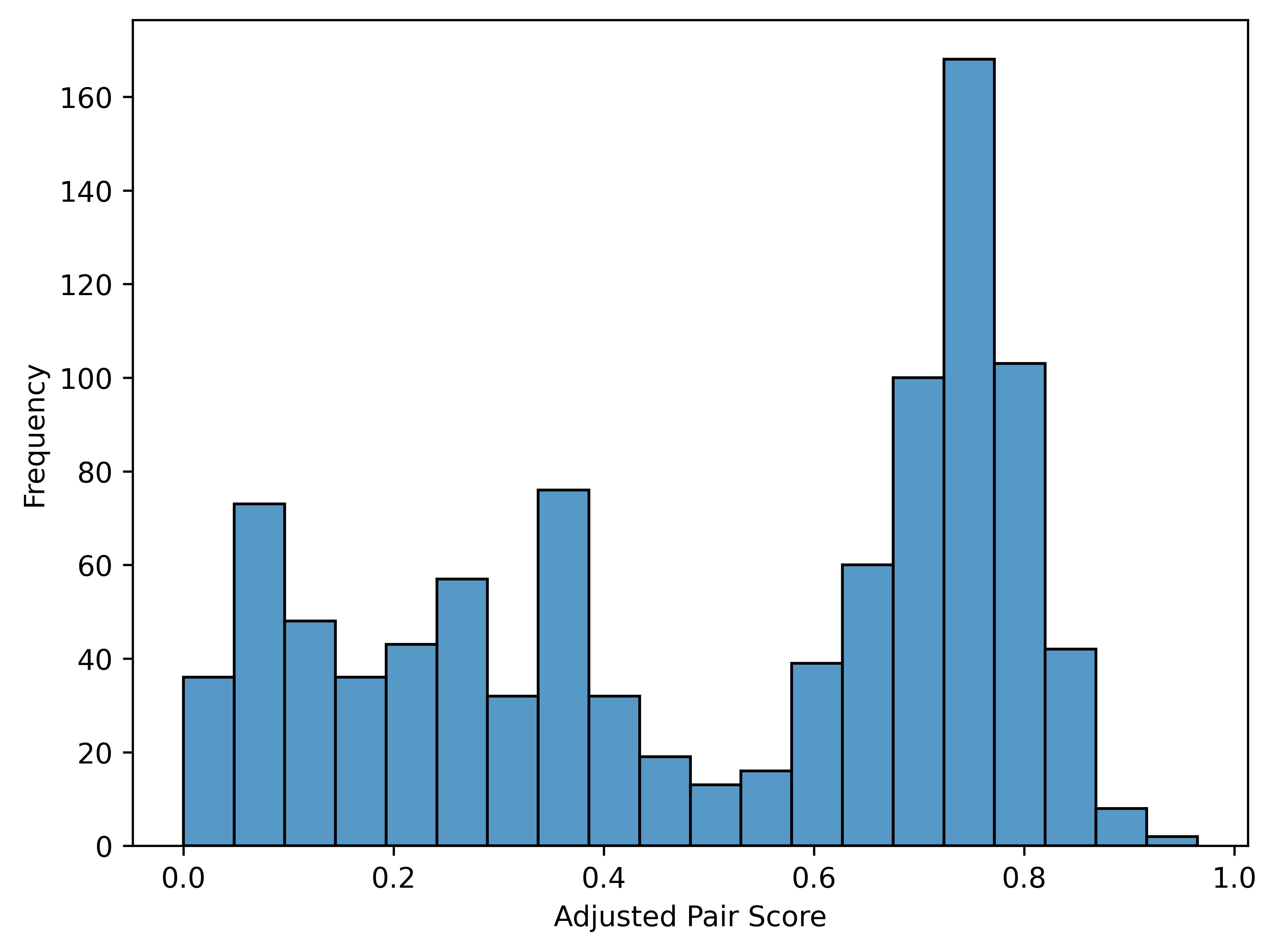}
    \caption{Histogram of Adjusted Pair Score for Case Study 2}
    \label{fig:judges_hist}
\end{figure}

In this section, we provide an analysis of the  adjusted pair score for Case Study 2. For this case study, the adjusted pair score has a mean of 0.500, a standard deviation of 0.269, and a correlation of -0.023 with the outcome of interest.  Figure \ref{fig:judges_hist} gives the distribution of the adjusted pair score. 
Table \ref{tab:gw_corr} displays the correlation of each covariate with our adjusted pair score. We see that the correlations are highest for the prior history variables, matching our assessment that the LLM bases its predictions strongly on these variables. Age is also slightly correlated, but the correlation is much smaller than the prior history variables.

\begin{table}[ht]
\centering
\small
\caption{Correlation of Covariates for Case Study 2}
\begin{tabular}{l c }
\hline
Covariate & Correlation with Adjusted Pair Score \\
\hline
\textit{Age } &  0.277 \\
\textit{Gender: Female } &  -0.227  \\
\textit{Race: Black } &  -0.110  \\
\textit{Prior History} & \\
\quad Prior Arrest &  0.644  \\
\quad Prior Drug Arrest &  0.668 \\
\quad Prior Felony Arrest &  0.795 \\
\quad Prior Drug Felony Arrest &  0.751 \\
\quad Prior Conviction &  0.806 \\
\quad Prior Drug Conviction &  0.730 \\
\quad Prior Felony Conviction &  0.914 \\
\quad Prior Drug Felony Conviction &  0.759 \\
\textit{Charge} & \\
\quad Possession with intent to Distribute & -0.022 \\
\quad Distribution &  0.052 \\
\quad{Additional Non Drug Charge} & -0.056   \\ 
\textit{Drugs Involved} & \\
\quad Cocaine &  -0.087  \\
\quad Marijuana & -0.058  \\ 
\quad Crack & -0.056  \\
\quad Heroin & 0.255  \\
\quad PCP &  -0.044  \\
\quad Other Drug &  -0.056  \\[0.5em]
\hline
\label{tab:gw_corr}
\end{tabular}

\vspace{0.5em}
\label{tab:gw_cov_corr}
\end{table}

\FloatBarrier
\subsection{Additional Results for Case Study 2}
\FloatBarrier

\begin{figure}[htbp]
    \centering
    \includegraphics[width=0.5\textwidth]{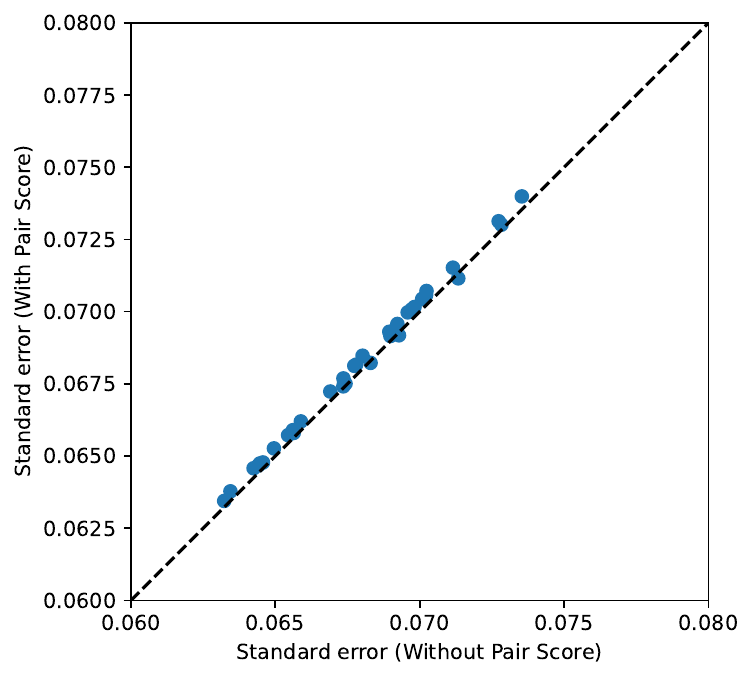}
    \caption{Standard Error with Pair Score vs. Standard Error Without Pair Score}
    \label{fig:gw_scatter}
\end{figure}

Table \ref{tab:gw_regress_results} compares a logistic model with and without the adjusted pair score. As discussed in the paper, the adjusted pair score is not statistically significant. Table \ref{tab:gw_var_results} gives the standard errors of the estimators with and without the adjusted pair score and Figure \ref{fig:gw_scatter} compares these in a scatterplot. The standard errors are very similar with and without the pair score. 

\begin{table}[ht]
\centering
\small
\caption{Logistic Regression Results for Case Study 2}
\begin{tabular}{lccc ccc}
\hline
& \multicolumn{3}{c}{Baseline Model} & \multicolumn{3}{c}{LLM Model} \\
\cline{2-4} \cline{5-7}
Variable & Estimate & SE & p & Estimate & SE & p \\
\hline
Intercept & 1.552 & 0.494   & 0.002   & 1.576 & 0.501  & 0.002  \\
\textbf{Adjusted Pair Score} &  &  &  & \textbf{-0.300} & \textbf{1.442} & \textbf{0.792} \\
OOB Predictions & 1.276 & 0.369  & 0.001  & 1.253  & 0.378  & 0.001  \\
Age & -0.051 & 0.008 & 0.000  & -0.051  & 0.008  & 0.000  \\
Female & -0.212 & 0.222  & 0.338  & -0.239 & 0.244  & 0.327  \\
Black & -1.128 & 0.476 & 0.018  & -1.135  & 0.477  & 0.017  \\
Prior Arrest & -0.433 & 0.317 & 0.172  & -0.395  & 0.348  & 0.256  \\
Prior Drug Arrest & 0.054 & 0.286  & 0.176  & 0.078  & 0.300 & 0.795 \\
Prior Felony Arrest & 0.578 & 0.278 & 0.038 & 0.616 &  0.313 & 0.049  \\
Prior Felony Drug Arrest &  -0.576 & 0.299 & 0.054  & -0.574  & 0.299  & 0.054  \\
Prior Conviction & 0.213 &  0.314 & 0.497 & 0.229  & 0.319 & 0.473 \\
Prior Drug Conviction & 0.273 & 0.317  & 0.389 & 0.289 & 0.323 & 0.371 \\
Prior Felony Conviction & -0.452 & 0.310  & 0.144  & -0.333  & 0.550  & 0.545  \\
Prior Drug Felony Conviction & 0.386 & 0.336  & 0.251  & 0.383  & 0.337 & 0.256 \\
Possession with intent & 0.138 & 0.254 & 0.586 & 0.141  & 0.255 &  0.578 \\
Distribution & 0.193 & 0.259 & 0.456  & 0.191  & 0.259  & 0.460  \\
Additional non-drug charge & 0.033 & 0.202  & 0.869 & 0.054 & 0.216 & 0.804  \\
Drug: Cocaine & -0.069 & 0.242  & 0.774  & -0.044  & 0.261  & 0.867  \\
Drug: Marijuana & 0.433 & 0.239 & 0.070  & 0.422  & 0.243  & 0.083 \\
Drug: Crack & 0.127 & 0.271 & 0.640  & 0.161  & 0.301  & 0.592 \\
Drug: Heroin & 0.360 & 0.259  & 0.165 &  0.392 & 0.287 & 0.172 \\
Drug: PCP & 0.573 &  0.387 & 0.139 & 0.594 & 0.396 & 0.133  \\
Drug: Other Drug & -0.432 & 0.458 & 0.345  & -0.416  & 0.462  & 0.368 \\
Judge 1 & -0.878 & 0.289 & 0.002  & -0.886 &  0.290 & 0.002 \\
Judge 2 & -0.892 & 0.293 & 0.002 & -0.897 &  0.294 &  0.002\\
Judge 3 & -0.387 & 0.289 & 0.180 & -0.394 & 0.290 &  0.174 \\
Judge 4 & -0.801 & 0.284 & 0.005 & -0.804 & 0.284 &  0.005 \\
Judge 5 & -0.547 & 0.291 & 0.060 & -0.552 & 0.291 &  0.058 \\
Judge 6 & -0.436 &  0.305 & 0.153  & -0.440 & 0.306  & 0.150 \\
Judge 7 & -0.826 &  0.295 & 0.005 &  -0.831 & 0.296 & 0.005 \\
Judge 8 & -0.937 & 0.299 & 0.002 &  -0.941 & 0.299 &  0.002\\
\hline
\end{tabular}
\label{tab:gw_regress_results}
\end{table}

\FloatBarrier
\begin{table}[ht]
\centering
\renewcommand{\arraystretch}{1.3}
\centering
\small
\caption{Estimator Standard Errors for All Pairwise Comparisons, Without Pair Score vs. With Pair Score}
\begin{tabular}{cccccccccc}
\toprule
 & Judge 1 & Judge 2 & Judge 3 & Judge 4 & Judge 5 & Judge 6 & Judge 7 & Judge 8 & Judge 9 \\
\midrule
Judge 1 & -- &
\makecell{0.068 \\ \textcolor{gray}{0.068}} &
\makecell{0.066\\\textcolor{gray}{0.066}} &
\makecell{0.064\\\textcolor{gray}{0.065}} &
\makecell{0.066\\\textcolor{gray}{0.066}} &
\makecell{0.071\\\textcolor{gray}{0.072}} &
\makecell{0.067\\\textcolor{gray}{0.068}} &
\makecell{0.069\\\textcolor{gray}{0.069}} &
\makecell{0.066\\\textcolor{gray}{0.066}} \\

Judge 2 &  &
-- &
\makecell{0.067\\\textcolor{gray}{0.068}} &
\makecell{0.065\\\textcolor{gray}{0.065}} &
\makecell{0.065\\\textcolor{gray}{0.065}} &
\makecell{0.073\\\textcolor{gray}{0.073}} &
\makecell{0.070\\\textcolor{gray}{0.070}} &
\makecell{0.071\\\textcolor{gray}{0.071}} &
\makecell{0.067\\\textcolor{gray}{0.068}} \\

Judge 3 &  &  &
-- &
\makecell{0.063\\\textcolor{gray}{0.063}} &
\makecell{0.065\\\textcolor{gray}{0.066}} &
\makecell{0.070\\\textcolor{gray}{0.070}} &
\makecell{0.068\\\textcolor{gray}{0.068}} &
\makecell{0.069\\\textcolor{gray}{0.069}} &
\makecell{0.067\\\textcolor{gray}{0.067}} \\

Judge 4 &  &  &  &
-- &
\makecell{0.064\\\textcolor{gray}{0.065}} &
\makecell{0.068\\\textcolor{gray}{0.068}} &
\makecell{0.067\\\textcolor{gray}{0.067}} &
\makecell{0.069\\\textcolor{gray}{0.069}} &
\makecell{0.063\\\textcolor{gray}{0.063}} \\

Judge 5 &  &  &  &  &
-- &
\makecell{0.070\\\textcolor{gray}{0.070}} &
\makecell{0.068\\\textcolor{gray}{0.068}} &
\makecell{0.069\\\textcolor{gray}{0.070}} &
\makecell{0.064\\\textcolor{gray}{0.065}} \\

Judge 6 &  &  &  &  &  &
-- &
\makecell{0.073\\\textcolor{gray}{0.073}} &
\makecell{0.074\\\textcolor{gray}{0.074}} &
\makecell{0.070\\\textcolor{gray}{0.071}} \\

Judge 7 &  &  &  &  &  &  &
-- &
\makecell{0.070\\\textcolor{gray}{0.070}} &
\makecell{0.069\\\textcolor{gray}{0.070}} \\

Judge 8 &  &  &  &  &  &  &  &
-- &
\makecell{0.070\\\textcolor{gray}{0.071}} \\

Judge 9 &  &  &  &  &  &  &  &  &
-- \\

\bottomrule
\end{tabular}
\vspace{0.5em}

\footnotesize{Each upper-triangular cell reports the standard error of the estimator without the adjusted pair score (top) and the standard error of the estimator with the adjusted pair score (\textcolor{gray}{bottom}). Standard errors are calculated assuming an equal chance of assignment to either of the  two judge calendars.}
\label{tab:gw_var_results}
\end{table}

\FloatBarrier

\FloatBarrier
\clearpage
\section{Supplementary Material for Case Study 3}
\label{sec:cta_appendix}

\subsection{Summary Statistics for Case Study 3 }
\FloatBarrier

Here, we give basic summary statistics for Case Study 3. Table \ref{cta_cov} displays the sample size and percentage of observations in each category for the categorical variables. Missing values are denoted by an ``Unknown'' category.

\begin{table}[ht!]
\centering
\small
\caption{Summary statistics of covariates for Case Study 3}
\label{tab:covariates2}
\begin{tabular}{l c c }
\hline
Covariate & n & Percentage \\
\hline

\textbf{Categorical covariates} & \\
\textit{Sex} & \\
\quad Female & 9396 & 49  \\
\quad Male & 9092 & 47 \\
\quad Unknown & 565 & 3 \\[0.5em]
\textit{Race} & \\
\quad White non-Hispanic & 9064 & 48 \\
\quad Black non-Hispanic & 4606 & 24 \\
\quad Hispanic & 3535 & 19 \\
\quad Asian/Pacific Islander & 481 & 3 \\
\quad Other Race/Multi-Racial & 112 & 1 \\
\quad American Indian/Alaskan Native & 35 & 0.5 \\
\quad Unknown & 1220 & 6 \\[0.5em]
\textit{Grade Level} & \\
\quad Middle school & 5930 & 31 \\
\quad High School & 13123 & 69 \\[0.5em]
\textit{State} & \\
\quad Texas & 5758 & 30 \\
\quad Kentucky & 5209 & 27 \\
\quad Louisiana & 2986 & 16 \\
\quad Michigan & 2449 & 13 \\
\quad Connecticut & 1027 & 5 \\
\quad Alabama & 849 & 4 \\
\quad New Jersey & 775 & 4 \\[0.5em]
\textit{English as a Second Language Education} & \\
\quad Yes & 290 & 2 \\
\quad No & 17182 & 90 \\
\quad Unknown & 1581 & 8 \\[0.5em]
\textit{Gifted Education} & \\
\quad Yes & 2351 & 12 \\
\quad No & 16492 & 87 \\
\quad Unknown & 210 & 1 \\[0.5em]
\textit{Special Education} & \\
\quad Yes & 1029 & 5 \\
\quad No & 17814 & 87 \\
\quad Unknown & 210 & 1 \\[0.5em]
\textit{Free and Reduced Lunch} & \\
\quad Yes & 8755 & 46 \\
\quad No & 6000 & 31 \\
\quad Unknown & 4298 & 23 \\[0.5em]
\hline
\end{tabular}

\vspace{0.5em}
\footnotesize{
N = 19,053. 
}
\label{cta_cov}
\end{table}

\FloatBarrier
\subsection{Analysis of Adjusted Pair Score for Case Study 3}
\FloatBarrier

\begin{figure}[htbp]
    \centering
    \includegraphics[width=0.8\textwidth]{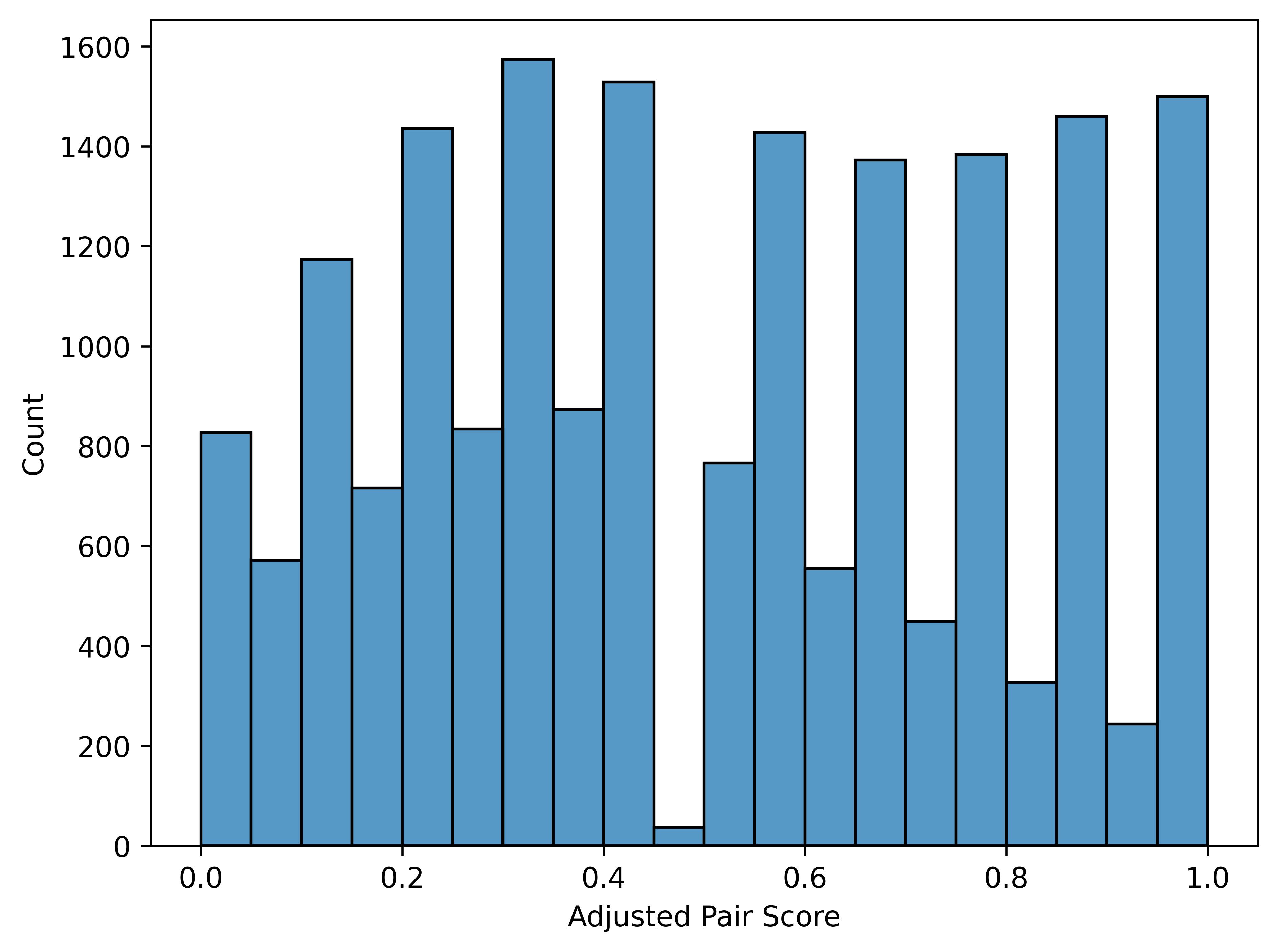}
    \caption{Histogram of Adjusted Pair Score for Case Study 3}
    \label{fig:cta_hist}
\end{figure}

For this case study, the adjusted pair score has a mean of 0.50, a standard deviation of 0.294, and a correlation of 0.175 with the outcome of interest. Figure \ref{fig:cta_hist} displays a histogram of the adjusted pair score values. Table \ref{tab:cta_cov2} displays the correlation of each covariate with our adjusted pair score. The correlation is highest for the pretest score; all other correlations are very small. 

\begin{table}[ht]
\centering
\small
\caption{Correlation of Covariates for Case Study 3 (UPDATED???)}
\begin{tabular}{l c}
\hline
Covariate & Correlation with Adjusted Pair Score \\
\hline
\textit{Pretest Score } & 0.481  \\
\textit{Sex: Male} & -0.021  \\
\textit{Race} & \\
\quad White non-Hispanic & 0.149  \\
\quad Black non-Hispanic & -0.099  \\
\quad Hispanic & -0.070 \\
\quad Asian/Pacific Islander &  -0.016 \\
\quad Other Race/Multi-Racial &  -0.006 \\
\quad American Indian/Alaskan Native & -0.005  \\
\textit{Grade Level: Middle School} & -0.046  \\
\textit{State} & \\
\quad Texas & -0.198  \\
\quad Kentucky & 0.139 \\
\quad Louisiana & 0.085  \\
\quad Michigan & -0.047  \\
\quad Connecticut & 0.027  \\
\quad Alabama & 0.053 \\
\quad New Jersey & -0.019  \\[0.5em]
\textit{Special Education} & -0.134 \\
\textit{Gifted Education} & 0.070  \\
\textit{English as a Second Language Education} & -0.081 \\
\textit{Free and Reduced Lunch} & -0.002  \\[0.5em]
\hline
\end{tabular}

\vspace{0.5em}
\label{tab:cta_cov2}
\end{table}

\FloatBarrier
\subsection{Additional Results for Case Study 3}
\FloatBarrier

Table \ref{tab:cta_results} compares a regression model with and without the adjusted pair score. As discussed in the paper, the adjusted pair score is statistically significant.

\begin{table}[ht!]
\centering
\small
\caption{Regression Results for Case Study 3}
\begin{tabular}{lccc ccc}
\hline
& \multicolumn{3}{c}{Baseline Model} & \multicolumn{3}{c}{LLM Model} \\
\cline{2-4} \cline{5-7}
Variable & Estimate & SE & p & Estimate & SE & p \\
\hline
Intercept & -0.290 & 0.122 & 0.019 & -0.368 & 0.123 & 0.003\\
\textbf{Adjusted Pair Score} & & & & \textbf{0.153} & \textbf{0.028} & \textbf{0.000}\\
OOB Predictions & 0.105 & 0.013 & 0.000 & 0.168 & 0.017 & 0.000 \\
Pretest Score & 0.380 & 0.009 & 0.000 & 0.315 & 0.015 & 0.000 \\
Pretest Score Unknown & -0.144 & 0.014 & 0.000 & -0.094 & 0.017 & 0.000 \\
Male & 0.019 & 0.009 & 0.037 & 0.020 & 0.009 & 0.026 \\
Gender Unknown & 0.026 & 0.037 & 0.488 & 0.021 &  0.037 & 0.567 \\
White non-Hispanic & 0.045 & 0.104 & 0.663 & 0.039 & 0.103 & 0.704 \\
Black non-Hispanic & -0.033 & 0.104 & 0.752 & -0.035 & 0.104 & 0.739 \\
Hispanic & -0.045 & 0.105 & 0.667 & -0.048  & 0.105 & 0.643 \\
Asian/Pacific Islander & 0.193 & 0.107 & 0.073 & 0.185 & 0.107 & 0.085 \\
Other Race/Multi Racial & 0.010 & 0.119 & 0.933 & 0.007 & 0.119 & 0.953 \\
Race Unknown & -0.027 & 0.107 & 0.803 & -0.030 & 0.107 & 0.782 \\
Grade Level: Middle School & 0.477 & 0.054 & 0.000 & 0.482 & 0.054 & 0.000 \\
English as a Second Language & 0.022 & 0.038 & 0.560 & 0.027 & 0.038 & 0.486 \\
ESL Unknown & 0.099 & 0.021 & 0.000 & 0.096 & 0.021 & 0.000 \\
Gifted Education & 0.106 & 0.017 & 0.000 & 0.101 & 0.017 & 0.000 \\
Special Education & -0.136 & 0.021 & 0.000 & -0.118 & 0.021 & 0.000 \\
Special Education Unknown & -0.220 & 0.062 & 0.000 & -0.208 & 0.062 & 0.001 \\
Free and Reduced Lunch & -0.010 &  0.012 & 0.396 & -0.011 & 0.012 & 0.358 \\
Free and Reduced Lunch Unknown & -0.076 & 0.017 & 0.000 & -0.077 & 0.017 & 0.000 \\
Treatment & 0.005 & 0.054 & 0.933  & 0.007  & 0.054 & 0.902 \\
\hline
\end{tabular}
\label{tab:cta_results}
\vspace{0.5em}

\footnotesize{Regression models are fit as multilevel linear models using the `lmerTest' package in R for p-values. Models have three levels: individuals are clustered within schools, and schools are clustered within states. The covariate capturing whether a student's gifted education status is unknown is dropped as it correlates perfectly with `Special Education Unknown'.
}
\end{table}

\FloatBarrier
\subsection{Comparison to Ranking Based Indicator}
\FloatBarrier
A reviewer wondered if it was the ranking itself, based on pairwise comparisons, that is useful. Here, we do an experiment to investigate. For each pair, we replaced the LLM response with a comparison of the out of bag predictions from a random forest on the full dataset. Whichever student had the higher out of bag prediction was selected as the more likely student to score higher. We converted this into an adjusted pair score using the same method of processing as the LLM predictions. These random forest based adjusted pair scores would be equivalent to the LLM based adjusted pair scores if the LLM always picked the student with the higher out-of-bag prediction. Unlike the LLM based pair score, the random forest based adjusted pair scores—which are based on rankings—is not significant in the model. Table \ref{tab:cta_ranking} displays the model results with this ranking-based pair score.

\begin{table}[ht!]
\centering
\small
\caption{Regression Results with Random Forest Ranking-Based Pair Score for Case Study 3}
\begin{tabular}{lccc ccc}
\hline
& \multicolumn{3}{c}{Baseline Model} & \multicolumn{3}{c}{LLM Model} \\
\cline{2-4} \cline{5-7}
Variable & Estimate & SE & p & Estimate & SE & p \\
\hline
Intercept & -0.290 & 0.122 & 0.019 & -0.295 & 0.123 & 0.017\\
\textbf{Ranking Based Adjusted Pair Score} & & & & \textbf{0.011} & \textbf{0.014} & \textbf{0.428}\\
OOB Predictions & 0.105 & 0.013 & 0.000 & 0.105 & 0.013 & 0.000 \\
Pretest Score & 0.380 & 0.009 & 0.000 & 0.380 & 0.009 & 0.000 \\
Pretest Score Unknown & -0.144 & 0.014 & 0.000 & -0.144 & 0.014 & 0.000 \\
Male & 0.019 & 0.009 & 0.037 & 0.019 & 0.009 & 0.037 \\
Gender Unknown & 0.026 & 0.037 & 0.488 & 0.026 &  0.037 & 0.483 \\
White non-Hispanic & 0.045 & 0.104 & 0.663 & 0.045 & 0.104 & 0.664 \\
Black non-Hispanic & -0.033 & 0.104 & 0.752 & -0.033 & 0.104 & 0.750 \\
Hispanic & -0.045 & 0.105 & 0.667 & -0.045  & 0.105 & 0.665 \\
Asian/Pacific Islander & 0.193 & 0.107 & 0.073 & 0.193 & 0.107 & 0.073 \\
Other Race/Multi Racial & 0.010 & 0.119 & 0.933 & 0.010 & 0.119 & 0.935 \\
Race Unknown & -0.027 & 0.107 & 0.803 & -0.027 & 0.107 & 0.800 \\
Grade Level: Middle School & 0.477 & 0.054 & 0.000 & 0.477 & 0.054 & 0.000 \\
English as a Second Language & 0.022 & 0.038 & 0.560 & 0.022 & 0.038 & 0.560 \\
ESL Unknown & 0.099 & 0.021 & 0.000 & 0.099 & 0.021 & 0.000 \\
Gifted Education & 0.106 & 0.017 & 0.000 & 0.106 & 0.017 & 0.000 \\
Special Education & -0.136 & 0.021 & 0.000 & -0.136 & 0.021 & 0.000 \\
Special Education Unknown & -0.220 & 0.062 & 0.000 & -0.220 & 0.062 & 0.000 \\
Free and Reduced Lunch & -0.010 &  0.012 & 0.396 & -0.010 & 0.012 & 0.401 \\
Free and Reduced Lunch Unknown & -0.076 & 0.017 & 0.000 & -0.076 & 0.017 & 0.000 \\
Treatment & 0.005 & 0.054 & 0.993  & 0.005  & 0.054 & 0.933 \\
\hline
\end{tabular}
\label{tab:cta_ranking}
\vspace{0.5em}

\footnotesize{Regression models are fit as multilevel linear models using the `lmerTest' package in R for p-values. Models have three levels: individuals are clustered within schools, and schools are clustered within states. The covariate capturing whether a student's gifted education status is unknown is dropped as it correlates perfectly with `Special Education Unknown'.
}
\end{table}

\FloatBarrier
\section{LLM Example Prompts}
\label{sec:prompt_appendix}

All LLM prompts incorporate a significant amount of observation-specific information. This information is presented naturally in sentence form. As a result, the content of our prompts changes slightly for different pairs of observations. Here, we give example prompts for each case study. For more details on all the forms these prompts may take, see the code at \href{https://github.com/jaylinlowe/llm-rct}{github.com/jaylinlowe/llm-rct}.

\subsection{Prompts for Case Study 1}

For this case study, we give an example template of the LLM prompts. The LLM was given two queries for each ordered pair. The first asks which paper will receive more citations, while the second asks which paper best exhibits each of 10 qualities.

The template for the first query is: 

\begin{tcolorbox}[title=Case Study 1 Prompt Example, colback=gray!5, colframe=gray!50, before skip=5pt, after skip=5pt]
\textbf{Prompt:} I have information on two papers published in Science. Using only the information I give you, which paper do you believe will have more citations? Paper 1 was published in 2007 and was titled ``\textit{Title}''. Paper's 1 abstract is ``\textit{Abstract}''. Paper 2 was published in 2007 and was titled ``\textit{Title}''. Paper's 2 abstract is ``\textit{Abstract}''. Please respond either `Paper 1' or `Paper 2' with no additional text.
\end{tcolorbox}
\vspace{0.5 cm}

The template for the second query is:
\begin{tcolorbox}[title=Case Study 1 Prompt Example, colback=gray!5, colframe=gray!50, before skip=5pt, after skip=5pt]
\textbf{Prompt:} I have information on two papers published in Science. Paper 1 was published in 2007 and was titled ``\textit{Title}''. Paper's 1 abstract is ``\textit{Abstract}''. Paper 2 was published in 2007 and was titled ``\textit{Title}''. Paper's 2 abstract is ``\textit{Abstract}''. I would like you to decide which paper best exhibits each one of these 11 qualities. The qualities are: topic novelty, topic popularity, title catchiness, generalizability, writing quality, impact of results, subfield popularity, technicality, meaningful contributions, journal fit for Science, and applicability. For each quality, please respond 'Paper 1' or 'Paper 2' depending on which one you think best exhibits the quality. Your answer should consist of 11 responses in order, each separated by commas. 
\end{tcolorbox}
\vspace{0.5 cm}

Note that the response to ``journal fit for Science'' was not used, due to an mistake in the code that did not update the journal for each query.

\subsection{Prompts for Case Study 2}
This case study has eight binary overlapping covariates: whether the defendants have a prior arrest, prior felony arrest, prior drug related arrest, prior felony drug related arrest, prior conviction, prior felony conviction, prior drug related conviction, and prior felony drug related conviction. Instead of writing a separate sentence for each binary covariate, we employ a series of decision rules to write 1-2 sentences for each defendant’s history. For example, the history of a defendant who has a felony conviction, a drug conviction, a felony drug arrest, but no felony drug conviction is given as: ``In the past, they were arrested and convicted on two separate charges, one of which was a felony and one of which was drug related. In addition, they have been arrested for a felony drug charge in the past, but they weren’t convicted of it.” This person must also have a prior arrest, a prior drug arrest, and a prior felony arrest, but those additional details are incorporated into the history without additional sentences. As the prompt for this case study is more complex and differs substantially by observation, we provide an example rather than a template. The exact prompt options can be found in the code at \href{https://github.com/jaylinlowe/llm-rct}{github.com/jaylinlowe/llm-rct}.

An example prompt is:

\begin{tcolorbox}[title=Case Study 2 Prompt Example, colback=gray!5, colframe=gray!50, before skip=5pt, after skip=5pt]
\textbf{Prompt:} I have information on two specific defendants who were just arrested in Washington, DC on drug related charges. Person 1 is male, 48 years old, and was just arrested and charged with distribution of marijuana and heroin plus at least one other drug not in one of these categories. They have no prior arrests. Person 2 is female, 60 years old, and was just arrested and charged with possession and intent to distribute a drug other than marijuana, cocaine, crack, heroin, or PCP. This may include prescription drugs or ecstasy. They also were charged with at least one non-drug related charge. In the past, they were arrested and convicted on a felony charge. Using your knowledge of criminal records and likelihood of reoffending, which of these two people is more likely to be arrested again, on any criminal charge, in the next four years? Please respond either `Person 1' or `Person 2' with no additional text.
\end{tcolorbox}
\vspace{0.5 cm}

\subsection{Prompts for Case Study 3}

An example prompt is: 
\begin{tcolorbox}[title=Case Study 3 Example Prompt, colback=gray!5, colframe=gray!50, before skip=5pt, after skip=5pt]
\textbf{Prompt:} I am going to give you information about two students. Both just took an algebra class at their school. Using only the information I give you, which student do you think will score higher on an algebra proficiency test administered at the end of the class? Student 1 is a middle school student in Texas. They are male, White non-Hispanic, and whether they receive free or reduced lunch is unknown. They are not in special education, are not in gifted education, and are not in ESL education. Student 1's standardized score on an algebra readiness exam administered before the course was 1.2. Student 2 is a high school student in Michigan. They are female, American Indian or Alaskan Native, and do not receive free or reduced lunch. They are not in special education, are in gifted education, and are not in ESL education. Student 2's standardized score on an algebra readiness exam administered before the course was 0.9. Please respond either 'Student 1' or 'Student 2' with no additional text.
\end{tcolorbox}
\vspace{0.5 cm}

\FloatBarrier

\newpage
\clearpage
\printbibliography

@article{robins_aipw,
 URL = {http://www.jstor.org/stable/2290910},
 author = {James M. Robins and Andrea Rotnitzky and Lue Ping Zhao},
 journal = {Journal of the American Statistical Association},
 number = {427},
 pages = {846--866},
 publisher = {[American Statistical Association, Taylor & Francis, Ltd.]},
 title = {Estimation of Regression Coefficients When Some Regressors Are Not Always Observed},
 volume = {89},
 year = {1994}
}

@misc{mann2026generalframeworkdesignbasedtreatment,
      title={A General Framework for Design-Based Treatment Effect Estimation in Paired Cluster-Randomized Experiments}, 
      author={Charlotte Z. Mann and Adam C. Sales and Johann A. Gagnon-Bartsch},
      year={2026},
      eprint={2407.01765},
      archivePrefix={arXiv},
      primaryClass={stat.ME},
      url={https://arxiv.org/abs/2407.01765}, 
}

@article{breiman_rf_2001,
	title = {Random Forests},
    volume = {45}, 
	doi = {10.1023/A:1010933404324},
	journal = {Machine Learning},
	author = {Breiman, L},
	month = oct,
	year = {2001},
	pages = {5-32}
}

@article{fernandez_classifiers_2014,
	title = {Do we Need Hundreds of Classifiers to Solve Real World
Classification Problems?},
    volume = {15}, 
	issn = {1532-4435},
	journal = {Journal of Machine Learning Research},
	author = {Fernández-Delgado, Manuel and Cernadas, Eva and Barro, Senén and Amorim, Dinani},
	month = jan,
	year = {2014},
	pages = {3133-3181}
}

@article{horvitz_generalization_1952,
	title = {A {Generalization} of {Sampling} {Without} {Replacement} from a {Finite} {Universe}},
	volume = {47},
	%issn = {0162-1459, 1537-274X},
	url = {https://www.tandfonline.com/doi/full/10.1080/01621459.1952.10483446},
	doi = {10.1080/01621459.1952.10483446},
	language = {en},
	number = {260},
	journal = {Journal of the American Statistical Association},
	author = {Horvitz, D. G. and Thompson, D. J.},
	month = dec,
	year = {1952},
	pages = {663--685},
	file = {Horvitz and Thompson - 1952 - A Generalization of Sampling Without Replacement f.pdf:/Users/jaylincl/Zotero/storage/PEQTYRLS/Horvitz and Thompson - 1952 - A Generalization of Sampling Without Replacement f.pdf:application/pdf},
}

@article{oaxaca_male-female_1973,
	title = {Male-{Female} {Wage} {Differentials} in {Urban} {Labor} {Markets}},
	volume = {14},
	%issn = {0020-6598},
	url = {https://www.jstor.org/stable/2525981},
	doi = {10.2307/2525981},
	number = {3},
	journal = {International Economic Review},
	author = {Oaxaca, Ronald},
	year = {1973},
    publisher = {Wiley for Economics Department of the University of Pennsylvania and Institute of Social and Economic Research, Osaka University}, 
	pages = {693--709},
	file = {Full Text PDF:/Users/jaylincl/Zotero/storage/MD5LCIKS/Oaxaca - 1973 - Male-Female Wage Differentials in Urban Labor Mark.pdf:application/pdf},
}

@article{blinder_wage_1973,
	title = {Wage {Discrimination}: {Reduced} {Form} and {Structural} {Estimates}},
	volume = {8},
	%issn = {0022-166X},
	shorttitle = {Wage {Discrimination}},
	url = {https://www.jstor.org/stable/144855},
	doi = {10.2307/144855},
	abstract = {Regressions explaining the wage rates of white males, black males, and white females are used to analyze the white-black wage differential among men and the male-female wage differential among whites. A distinction is drawn between reduced form and structural wage equations, and both are estimated. They are shown to have very different implications for analyzing the white-black and male-female wage differentials. When the two sets of estimates are synthesized, they jointly imply that 70 percent of the overall race differential and 100 percent of the overall sex differential are ultimately attributable to discrimination of various sorts.},
	number = {4},
	journal = {The Journal of Human Resources},
	author = {Blinder, Alan S.},
	year = {1973},
    publisher = {University of Wisconsin Press, Board of Regents of the University of Wisconsin System},
	pages = {436--455},
	file = {Full Text PDF:/Users/jaylincl/Zotero/storage/8DZGI8FD/Blinder - 1973 - Wage Discrimination Reduced Form and Structural E.pdf:application/pdf},
}

@article{kurlychek_scarlet_2006,
	title = {Scarlet {Letters} and {Recidivism}: {Does} an {Old} {Criminal} {Record} {Predict} {Future} {Offending}?},
	volume = {5},
	%issn = {1745-9133},
	shorttitle = {Scarlet {Letters} and {Recidivism}},
	url = {https://onlinelibrary.wiley.com/doi/abs/10.1111/j.1745-9133.2006.00397.x},
	doi = {10.1111/j.1745-9133.2006.00397.x},
	abstract = {Research Summary: This research explores the issue of old prior records and their ability to predict future offending. In particular, we are interested in the question of whether, after a given period of time, the risk of recidivism for a person who has been arrested in the distant past is ever indistinguishable from that of a population of persons with no prior arrests. Two well-documented empirical facts guide our investigation: (1) Individuals who have offended in the past are relatively more likely to offend in the future, and (2) the risk of recidivism declines as the time since the last criminal act increases. We find that immediately after an arrest, the knowledge of this prior record does significantly differentiate this population from a population of nonoffenders. However, these differences weaken dramatically and quickly over time so that the risk of new offenses among those who last offended six or seven years ago begins to approximate (but not match) the risk of new offenses among persons with no criminal record. Policy Implications: Individuals with official records of past offending behavior encounter a barrier when they try to obtain employment, even if a person's most recent offense occurred in the distant past. There are many reasons for such obstacles, but they are at least partially premised on the concern that individuals with arrest records—even from the distant past—are more likely to offend in the future than persons with no criminal history. Our analysis questions the logic of such practices and suggests that after a given period of remaining crime free, it may be prudent to wash away the brand of “offender” and open up more legitimate opportunities to this population.},
	language = {en},
	number = {3},
	journal = {Criminology \& Public Policy},
	author = {Kurlychek, Megan C. and Brame, Robert and Bushway, Shawn D.},
	year = {2006},
	keywords = {Collateral Consequences, Desistance, Recidivism},
	pages = {483--504},
	file = {Full Text PDF:/Users/jaylincl/Zotero/storage/UHLGQBSF/Kurlychek et al. - 2006 - Scarlet Letters and Recidivism Does an Old Crimin.pdf:application/pdf;Snapshot:/Users/jaylincl/Zotero/storage/AZ6T6RFG/j.1745-9133.2006.00397.html:text/html},
}

@article{gendreau_meta-analysis_1996,
	title = {A {Meta}-{Analysis} of the {Predictors} of {Adult} {Offender} {Recidivism}: {What} {Works}!},
	volume = {34},
	%issn = {1745-9125},
	shorttitle = {A {Meta}-{Analysis} of the {Predictors} of {Adult} {Offender} {Recidivism}},
	url = {https://onlinelibrary.wiley.com/doi/abs/10.1111/j.1745-9125.1996.tb01220.x},
	doi = {10.1111/j.1745-9125.1996.tb01220.x},
	abstract = {Meta-analytic techniques were used to determine which predictor domains and actuarial assessment instruments were the best predictors of adult offender recidivism. One hundred and thirty-one studies produced 1,141 correlations with recidivism. The strongest predictor domains were criminogenic needs, criminal history/history of antisocial behavior, social achievement, age/gender/race, and family factors. Less robust predictors included intellectual functioning, personal distress factors, and socioeconomic status in the family of origin. Dynamic predictor domains performed at least as well as the static domains. The LSI-R was identified as the most useful actuarial measure. Recommendations for developing sound assessment practices in corrections are provided.},
	language = {en},
	number = {4},
	journal = {Criminology},
	author = {Gendreau, Paul and Little, Tracy and Goggin, Claire},
	year = {1996},
	pages = {575--608},
	file = {Full Text PDF:/Users/jaylincl/Zotero/storage/GEKWBHIE/Gendreau et al. - 1996 - A Meta-Analysis of the Predictors of Adult Offende.pdf:application/pdf;Snapshot:/Users/jaylincl/Zotero/storage/C26VZ4DQ/j.1745-9125.1996.tb01220.html:text/html},
}

@article{freedman_regression_2008,
	title = {On regression adjustments to experimental data},
	volume = {40},
	%issn = {01968858},
	url = {https://linkinghub.elsevier.com/retrieve/pii/S019688580700005X},
	doi = {10.1016/j.aam.2006.12.003},
	abstract = {Regression adjustments are often made to experimental data. Since randomization does not justify the models, almost anything can happen. Here, we evaluate results using Neyman’s non-parametric model, where each subject has two potential responses, one if treated and the other if untreated. Only one of the two responses is observed. Regression estimates are generally biased, but the bias is small with large samples. Adjustment may improve precision, or make precision worse; standard errors computed according to usual procedures may overstate the precision, or understate, by quite large factors. Asymptotic expansions make these ideas more precise.},
	language = {en},
	number = {2},
	journal = {Advances in Applied Mathematics},
	author = {Freedman, David A.},
	month = feb,
	year = {2008},
	pages = {180--193},
	file = {Freedman - 2008 - On regression adjustments to experimental data.pdf:/Users/jaylincl/Zotero/storage/Y5HE3U2H/Freedman - 2008 - On regression adjustments to experimental data.pdf:application/pdf},
}

@book{fisher_statistical_1925,
	address = {Oliver and Boyd},
	series = {Statistical methods for research workers, 11th ed. rev},
	title = {Statistical methods for research workers, 11th ed. rev},
	publisher = {Edinburgh},
	author = {Fisher, R.A.},
	year = {1925},
}

@article{lin_agnostic_2013,
	title = {Agnostic notes on regression adjustments to experimental data: {Reexamining} {Freedman}'s critique},
	volume = {7},
	%issn = {1932-6157},
	shorttitle = {Agnostic notes on regression adjustments to experimental data},
	url = {http://arxiv.org/abs/1208.2301},
	doi = {10.1214/12-AOAS583},
	abstract = {Freedman [Adv. in Appl. Math. 40 (2008) 180-193; Ann. Appl. Stat. 2 (2008) 176-196] critiqued ordinary least squares regression adjustment of estimated treatment effects in randomized experiments, using Neyman's model for randomization inference. Contrary to conventional wisdom, he argued that adjustment can lead to worsened asymptotic precision, invalid measures of precision, and small-sample bias. This paper shows that in sufficiently large samples, those problems are either minor or easily fixed. OLS adjustment cannot hurt asymptotic precision when a full set of treatment-covariate interactions is included. Asymptotically valid confidence intervals can be constructed with the Huber-White sandwich standard error estimator. Checks on the asymptotic approximations are illustrated with data from Angrist, Lang, and Oreopoulos's [Am. Econ. J.: Appl. Econ. 1:1 (2009) 136--163] evaluation of strategies to improve college students' achievement. The strongest reasons to support Freedman's preference for unadjusted estimates are transparency and the dangers of specification search.},
	number = {1},
	journal = {The Annals of Applied Statistics},
	author = {Lin, Winston},
	month = mar,
	year = {2013},
	note = {arXiv:1208.2301 [stat]},
	keywords = {Statistics - Applications, Statistics - Methodology},
}

@article{davis_open_2011,
	title = {Open access, readership, citations: a randomized controlled trial of scientific journal publishing},
	volume = {25},
	%issn = {1530-6860},
	shorttitle = {Open access, readership, citations},
	doi = {10.1096/fj.11-183988},
	abstract = {Does free access to journal articles result in greater diffusion of scientific knowledge? Using a randomized controlled trial of open access publishing, involving 36 participating journals in the sciences, social sciences, and humanities, we report on the effects of free access on article downloads and citations. Articles placed in the open access condition (n=712) received significantly more downloads and reached a broader audience within the first year, yet were cited no more frequently, nor earlier, than subscription-access control articles (n=2533) within 3 yr. These results may be explained by social stratification, a process that concentrates scientific authors at a small number of elite research universities with excellent access to the scientific literature. The real beneficiaries of open access publishing may not be the research community but communities of practice that consume, but rarely contribute to, the corpus of literature.},
	language = {eng},
	number = {7},
	journal = {FASEB Journal},
	author = {Davis, Philip M.},
	month = jul,
	year = {2011},
	pmid = {21450907},
	keywords = {Access to Information, Biomedical Research, Humans, Information Dissemination, Information Storage and Retrieval, Internet, Publishing},
	pages = {2129--2134},
}

@misc{zheng_how_2024,
	title = {How {Reliable} are {LLMs} as {Knowledge} {Bases}? {Re}-thinking {Facutality} and {Consistency}},
	shorttitle = {How {Reliable} are {LLMs} as {Knowledge} {Bases}?},
	url = {http://arxiv.org/abs/2407.13578},
	doi = {10.48550/arXiv.2407.13578},
	publisher = {arXiv},
	author = {Zheng, Danna and Lapata, Mirella and Pan, Jeff Z.},
	month = dec,
	year = {2024},
	note = {arXiv:2407.13578 [cs]},
}

@inproceedings{mahaut_factual_2024,
	title = {Factual {Confidence} of {LLMs}: on {Reliability} and {Robustness} of {Current} {Estimators}},
	shorttitle = {Factual {Confidence} of {LLMs}},
	url = {http://arxiv.org/abs/2406.13415},
	doi = {10.18653/v1/2024.acl-long.250},
	booktitle = {Proceedings of the 62nd {Annual} {Meeting} of the {Association} for {Computational} {Linguistics} ({Volume} 1: {Long} {Papers})},
	author = {Mahaut, Matéo and Aina, Laura and Czarnowska, Paula and Hardalov, Momchil and Müller, Thomas and Màrquez, Lluís},
	year = {2024},
	note = {arXiv:2406.13415 [cs]},
	pages = {4554--4570},
}

@misc{ye_assessing_2023,
	title = {Assessing {Hidden} {Risks} of {LLMs}: {An} {Empirical} {Study} on {Robustness}, {Consistency}, and {Credibility}},
	shorttitle = {Assessing {Hidden} {Risks} of {LLMs}},
	url = {http://arxiv.org/abs/2305.10235},
	doi = {10.48550/arXiv.2305.10235},
	publisher = {arXiv},
	author = {Ye, Wentao and Ou, Mingfeng and Li, Tianyi and chen, Yipeng and Ma, Xuetao and Yanggong, Yifan and Wu, Sai and Fu, Jie and Chen, Gang and Wang, Haobo and Zhao, Junbo},
	month = aug,
	year = {2023},
	note = {arXiv:2305.10235 [cs]},
}

@article{majeed_reliability_2024,
	title = {Reliability {Issues} of {LLMs}: {ChatGPT} a {Case} {Study}},
	volume = {1},
	%issn = {2641-8819},
	shorttitle = {Reliability {Issues} of {LLMs}},
	url = {https://ieeexplore.ieee.org/document/10602758/},
	doi = {10.1109/MRL.2024.3420849},
	abstract = {ChatGPT is a groundbreaking artificial intelligence (AI) invention, and this technology will see tremendous growth per the IEEE Computer Society’s 2024 technology predictions report.1 According to the report, generative AI applications top the list, and this paradigm is predicted to experience most of the advancements in the coming years. ChatGPT, a generative AI product, has demonstrated its effectiveness in many ways (e.g., answering questions, summarizing text, generating computer code, fixing programming bugs, and generating synthetic data). Despite the many promising applications, ChatGPT cannot produce desirable results for many difficult and pragmatic tasks [1]. For example, the inaccuracy from ChatGPT answers related to the emotional text is significantly high, owing to limited amounts of data—or no available data—concerning these tasks [1]. Similarly, ChatGPT can be manipulated to generate fake content, which can be hard to distinguish from real content. There are two schools of thought in the AI community about ChatGPT technology.},
	number = {4},
	journal = {IEEE Reliability Magazine},
	author = {Majeed, Abdul and Hwang, Seong Oun},
	month = dec,
	year = {2024},
	keywords = {Artificial intelligence, Chatbots, Codes, Generative AI, Large language models, Reliability engineering, Servers, Training},
	pages = {36--46},
	file = {Full Text PDF:/Users/jaylincl/Zotero/storage/A3LVEDG4/Majeed and Hwang - 2024 - Reliability Issues of LLMs ChatGPT a Case Study.pdf:application/pdf},
}

@article{rubin_causal_2005,
	title = {Causal {Inference} {Using} {Potential} {Outcomes}: {Design}, {Modeling}, {Decisions}},
	volume = {100},
	%issn = {0162-1459},
	shorttitle = {Causal {Inference} {Using} {Potential} {Outcomes}},
	url = {https://www.jstor.org/stable/27590541},
	abstract = {Causal effects are defined as comparisons of potential outcomes under different treatments on a common set of units. Observed values of the potential outcomes are revealed by the assignment mechanism—a probabilistic model for the treatment each unit receives as a function of covariates and potential outcomes. Fisher made tremendous contributions to causal inference through his work on the design of randomized experiments, but the potential outcomes perspective applies to other complex experiments and nonrandomized studies as well. As noted by Kempthorne in his 1976 discussion of Savage's Fisher lecture, Fisher never bridged his work on experimental design and his work on parametric modeling, a bridge that appears nearly automatic with an appropriate view of the potential outcomes framework, where the potential outcomes and covariates are given a Bayesian distribution to complete the model specification. Also, this framework crisply separates scientific inference for causal effects and decisions based on such inference, a distinction evident in Fisher's discussion of tests of significance versus tests in an accept/reject framework. But Fisher never used the potential outcomes framework, originally proposed by Neyman in the context of randomized experiments, and as a result he provided generally flawed advice concerning the use of the analysis of covariance to adjust for posttreatment concomitants in randomized trials.},
	number = {469},
	journal = {Journal of the American Statistical Association},
	author = {Rubin, Donald B.},
	year = {2005},
    publisher = {American Statistical Association, Taylor & Francis, Ltd.}, 
	pages = {322--331},
	file = {Full Text PDF:/Users/jaylincl/Zotero/storage/UWGQSEMK/Rubin - 2005 - Causal Inference Using Potential Outcomes Design,.pdf:application/pdf},
}

@inproceedings{deng_improving_2013,
	address = {New York, NY, USA},
	series = {{WSDM} '13},
	title = {Improving the sensitivity of online controlled experiments by utilizing pre-experiment data},
	isbn = {978-1-4503-1869-3},
	url = {https://dl.acm.org/doi/10.1145/2433396.2433413},
	doi = {10.1145/2433396.2433413},
	booktitle = {Proceedings of the sixth {ACM} international conference on {Web} search and data mining},
	publisher = {Association for Computing Machinery},
	author = {Deng, Alex and Xu, Ya and Kohavi, Ron and Walker, Toby},
	month = feb,
	year = {2013},
	pages = {123--132},
}

@article{gui_combining_2024,
	title = {Combining {Observational} and {Experimental} {Data} to {Improve} {Efficiency} {Using} {Imperfect} {Instruments}},
	volume = {43},
	%issn = {0732-2399},
	url = {https://pubsonline.informs.org/doi/abs/10.1287/mksc.2020.0435},
	doi = {10.1287/mksc.2020.0435},
	abstract = {Randomized controlled trials generate experimental variation that can credibly identify causal effects, but often suffer from limited scale, whereas observational data sets are large, but often violate desired identification assumptions. To improve estimation efficiency, I propose a method that leverages imperfect instruments—pretreatment covariates that satisfy the relevance condition, but may violate the exclusion restriction. I show that these imperfect instruments can be used to derive moment restrictions that, in combination with the experimental data, improve estimation efficiency. I outline estimators for implementing this strategy and show that my methods can reduce variance by up to 50\%; therefore, only half of the experimental sample is required to attain the same statistical precision. I apply my method to a search-listing data set from Expedia that studies the causal effect of search rankings on clicks and show that the method can substantially improve the precision. History: Puneet Manchanda served as the senior editor. Supplemental Material: The online appendix and data files are available at https://doi.org/10.1287/mksc.2020.0435.},
	number = {2},
	journal = {Marketing Science},
	author = {Gui, George Z.},
	month = jan,
	year = {2024},
	keywords = {data fusion, econometrics, endogeneity, experimental design, measurement and inference},
	pages = {378--391},
	file = {Submitted Version:/Users/jaylincl/Zotero/storage/A54EZIM7/Gui - 2024 - Combining Observational and Experimental Data to I.pdf:application/pdf},
}

@article{rosenman_combining_2023,
	title = {Combining observational and experimental datasets using shrinkage estimators},
	volume = {79},
	%issn = {1541-0420},
	doi = {10.1111/biom.13827},
	abstract = {We consider the problem of combining data from observational and experimental sources to draw causal conclusions. To derive combined estimators with desirable properties, we extend results from the Stein shrinkage literature. Our contributions are threefold. First, we propose a generic procedure for deriving shrinkage estimators in this setting, making use of a generalized unbiased risk estimate. Second, we develop two new estimators, prove finite sample conditions under which they have lower risk than an estimator using only experimental data, and show that each achieves a notion of asymptotic optimality. Third, we draw connections between our approach and results in sensitivity analysis, including proposing a method for evaluating the feasibility of our estimators.},
	language = {eng},
	number = {4},
	journal = {Biometrics},
	author = {Rosenman, Evan and Basse, Guillaume and Owen, Art B. and Baiocchi, Mike},
	month = dec,
	year = {2023},
	pmid = {36629736},
	keywords = {causal inference, Causality, data fusion, Probability, sensitivity analysis, shrinkage},
	pages = {2961--2973},
	file = {Submitted Version:/Users/jaylincl/Zotero/storage/PA6378Z9/Rosenman et al. - 2023 - Combining observational and experimental datasets .pdf:application/pdf},
}

@article{rosenman_propensity_2022,
	title = {Propensity score methods for merging observational and experimental datasets},
	volume = {41},
	%issn = {1097-0258},
	doi = {10.1002/sim.9223},
	abstract = {We consider how to merge a limited amount of data from a randomized controlled trial (RCT) into a much larger set of data from an observational data base (ODB), to estimate an average causal treatment effect. Our methods are based on stratification. The strata are defined in terms of effect moderators as well as propensity scores estimated in the ODB. Data from the RCT are placed into the strata they would have occupied, had they been in the ODB instead. We assume that treatment differences are comparable in the two data sources. Our first "spiked-in" method simply inserts the RCT data into their corresponding ODB strata. We also consider a data-driven convex combination of the ODB and RCT treatment effect estimates within each stratum. Using the delta method and simulations, we identify a bias problem with the spiked-in estimator that is ameliorated by the convex combination estimator. We apply our methods to data from the Women's Health Initiative, a study of thousands of postmenopausal women which has both observational and experimental data on hormone therapy (HT). Using half of the RCT to define a gold standard, we find that a version of the spiked-in estimator yields lower-MSE estimates of the causal impact of HT on coronary heart disease than would be achieved using either a small RCT or the observational component on its own.},
	language = {eng},
	number = {1},
	journal = {Statistics in Medicine},
	author = {Rosenman, Evan and Owen, Art B. and Baiocchi, Mike and Banack, Hailey R.},
	month = jan,
	year = {2022},
	pmid = {34671998},
	keywords = {Bias, causal inference, Causality, Databases, Factual, external validity, Female, Humans, observational studies, Propensity Score, randomized controlled trials, Research Design},
	pages = {65--86},
	file = {Submitted Version:/Users/jaylincl/Zotero/storage/BUHNNBKF/Rosenman et al. - 2022 - Propensity score methods for merging observational.pdf:application/pdf},
}

@inproceedings{kallus_removing_2018,
	title = {Removing {Hidden} {Confounding} by {Experimental} {Grounding}},
	volume = {31},
	url = {https://papers.nips.cc/paper_files/paper/2018/hash/566f0ea4f6c2e947f36795c8f58ba901-Abstract.html},
	booktitle = {Advances in {Neural} {Information} {Processing} {Systems}},
	publisher = {Curran Associates, Inc.},
	author = {Kallus, Nathan and Puli, Aahlad Manas and Shalit, Uri},
	year = {2018},
}

@article{green_winik_2010,
	title = {Using {Random} {Judge} {Assignments} to {Estimate} the {Effects} of {Incarceration} and {Probation} on {Recidivism} {Among} {Drug} {Offenders}},
	volume = {48},
	%issn = {1745-9125},
	url = {https://onlinelibrary.wiley.com/doi/abs/10.1111/j.1745-9125.2010.00189.x},
	doi = {10.1111/j.1745-9125.2010.00189.x},
	language = {en},
	number = {2},
	journal = {Criminology},
	author = {Green, Donald P. and Winik, Daniel},
	year = {2010},
	keywords = {drug crime, natural experiments, recidivism, specific deterrence},
	pages = {357--387},
	file = {Full Text PDF:/Users/jaylincl/Zotero/storage/NHJJ6E62/Green and Winik - 2010 - Using Random Judge Assignments to Estimate the Eff.pdf:application/pdf;Snapshot:/Users/jaylincl/Zotero/storage/JBQ9VDEM/j.1745-9125.2010.00189.html:text/html},
}

@article{splawa-neyman_application_1990,
	title = {On the {Application} of {Probability} {Theory} to {Agricultural} {Experiments}. {Essay} on {Principles}. {Section} 9},
	volume = {5},
	%issn = {0883-4237},
	url = {https://www.jstor.org/stable/2245382},
	number = {4},
    translator  = {Jerzy and Dabrowska, D. M. and Speed, T. P.},
	journal = {Statistical Science},
	author = {Neyman, Jerzy},
	year = {1923},
    publisher = {Institute of Mathematical Statistics}, 
	pages = {465--472},
	file = {Full Text PDF:/Users/jaylincl/Zotero/storage/PB2JCZ2M/Splawa-Neyman et al. - 1990 - On the Application of Probability Theory to Agricu.pdf:application/pdf},
}

@article{aronow_class_2013,
	title = {A {Class} of {Unbiased} {Estimators} of the {Average} {Treatment} {Effect} in {Randomized} {Experiments}},
	volume = {1},
	%issn = {2193-3685, 2193-3677},
	url = {https://www.degruyter.com/document/doi/10.1515/jci-2012-0009/html},
	doi = {10.1515/jci-2012-0009},
	language = {en},
	number = {1},
	journal = {Journal of Causal Inference},
	author = {Aronow, Peter M. and Middleton, Joel A.},
	month = may,
	year = {2013},
	pages = {135--154},
	file = {Aronow and Middleton - 2013 - A Class of Unbiased Estimators of the Average Trea.pdf:/Users/jaylincl/Zotero/storage/KHLGYRHD/Aronow and Middleton - 2013 - A Class of Unbiased Estimators of the Average Trea.pdf:application/pdf},
}

@article{wager_high-dimensional_2016,
	title = {High-dimensional regression adjustments in randomized experiments},
	volume = {113},
	url = {https://www.pnas.org/doi/10.1073/pnas.1614732113},
	doi = {10.1073/pnas.1614732113},
	number = {45},
	journal = {Proceedings of the National Academy of Sciences},
	author = {Wager, Stefan and Du, Wenfei and Taylor, Jonathan and Tibshirani, Robert J.},
	month = nov,
	year = {2016},
    publisher = {Proceedings of the National Academy of Sciences}, 
	pages = {12673--12678},
}

@article{tsiatis_covariate_2008,
	title = {Covariate adjustment for two-sample treatment comparisons in randomized clinical trials: a principled yet flexible approach},
	volume = {27},
	%issn = {0277-6715},
	shorttitle = {Covariate adjustment for two-sample treatment comparisons in randomized clinical trials},
	doi = {10.1002/sim.3113},
	language = {eng},
	number = {23},
	journal = {Statistics in Medicine},
	author = {Tsiatis, Anastasios A. and Davidian, Marie and Zhang, Min and Lu, Xiaomin},
	month = oct,
	year = {2008},
	pmid = {17960577},
	pmcid = {PMC2562926},
	keywords = {Algorithms, Data Interpretation, Statistical, Humans, Randomized Controlled Trials as Topic, Sampling Studies, Statistics, Nonparametric, Treatment Outcome},
	pages = {4658--4677},
	file = {Accepted Version:/Users/jaylincl/Zotero/storage/E7E6C7PA/Tsiatis et al. - 2008 - Covariate adjustment for two-sample treatment comp.pdf:application/pdf},
}

@article{gagnon-bartsch_precise_2023-1,
	title = {Precise unbiased estimation in randomized experiments using auxiliary observational data},
	volume = {11},
	copyright = {De Gruyter expressly reserves the right to use all content for commercial text and data mining within the meaning of Section 44b of the German Copyright Act.},
	%issn = {2193-3685},
	url = {https://www.degruyter.com/document/doi/10.1515/jci-2022-0011/html?lang=en},
	doi = {10.1515/jci-2022-0011},
	abstract = {Randomized controlled trials (RCTs) admit unconfounded design-based inference – randomization largely justifies the assumptions underlying statistical effect estimates – but often have limited sample sizes. However, researchers may have access to big observational data on covariates and outcomes from RCT nonparticipants. For example, data from A/B tests conducted within an educational technology platform exist alongside historical observational data drawn from student logs. We outline a design-based approach to using such observational data for variance reduction in RCTs. First, we use the observational data to train a machine learning algorithm predicting potential outcomes using covariates and then use that algorithm to generate predictions for RCT participants. Then, we use those predictions, perhaps alongside other covariates, to adjust causal effect estimates with a flexible, design-based covariate-adjustment routine. In this way, there is no danger of biases from the observational data leaking into the experimental estimates, which are guaranteed to be exactly unbiased regardless of whether the machine learning models are “correct” in any sense or whether the observational samples closely resemble RCT samples. We demonstrate the method in analyzing 33 randomized A/B tests and show that it decreases standard errors relative to other estimators, sometimes substantially.},
	language = {en},
	number = {1},
	journal = {Journal of Causal Inference},
	author = {Gagnon-Bartsch, Johann A. and Sales, Adam C. and Wu, Edward and Botelho, Anthony F. and Erickson, John A. and Miratrix, Luke W. and Heffernan, Neil T.},
	month = jan,
	year = {2023},
        pages = {20220011},
    publisher = {De Gruyter}, 
	keywords = {A/B testing, data integration, education research},
	file = {Full Text PDF:/Users/jaylincl/Zotero/storage/IP8T3W9F/Gagnon-Bartsch et al. - 2023 - Precise unbiased estimation in randomized experime.pdf:application/pdf},
}

@article{pane_effectiveness_2014,
	title = {Effectiveness of {Cognitive} {Tutor} {Algebra} {I} at {Scale}},
	volume = {36},
	%issn = {0162-3737},
	url = {https://www.jstor.org/stable/43773458},
	abstract = {This article examines the effectiveness of a technology-based algebra curriculum in a wide variety of middle schools and high schools in seven states. Participating schools were matched into similar pairs and randomly assigned to either continue with the current algebra curriculum for 2 years or to adopt Cognitive Tutor Algebra I (CTAI), which uses a personalized, mastery-learning, blendedlearning approach. Schools assigned to implement CTAI did so under conditions similar to schools that independently adopt it. Analysis of posttest outcomes on an algebra proficiency exam finds no effects in the first year of implementation, but finds evidence in support of positive effects in the second year. The estimated effect is statistically significant for high schools but not for middle schools; in both cases, the magnitude is sufficient to improve the median student's performance by approximately eight percentile points.},
	number = {2},
	journal = {Educational Evaluation and Policy Analysis},
	author = {Pane, John F. and Griffin, Beth Ann and McCaffrey, Daniel F. and Karam, Rita},
	year = {2014},
    publisher = {Amercian Educational Research Association, Sage Publications, Inc.}, 
	pages = {127--144},
	file = {JSTOR Full Text PDF:/Users/jaylincl/Zotero/storage/LDEXSGRV/Pane et al. - 2014 - Effectiveness of Cognitive Tutor Algebra I at Scal.pdf:application/pdf},
}

\end{document}